\documentclass[11pt]{article}

\usepackage{enumerate, color, multirow}
\usepackage{amsbsy, amstext, amssymb, amsthm, amsmath,bm}
\usepackage{graphicx,booktabs,natbib}
\usepackage{algorithmic,algorithm}
\usepackage[margin=1in]{geometry}
\usepackage{array,stmaryrd}

\newtheorem{proposition}{{\bf Proposition}}
\newtheorem{lemma}{{\bf Lemma}}

\newtheorem{theorem}{{\bf Theorem}}

\newcommand{\vect}{\mathrm{vec}}

\newcommand{\real}[1]{\mathrm{I \! R} \mathit{^{#1}}}
\newcommand{\trans}{^{\mbox{\tiny {\sf T}}}}

\newcommand{\Abf}{{\bm A}}
\newcommand{\Bbf}{{\bm B}}
\newcommand{\Dbf}{{\bm D}}

\newcommand{\Hbf}{{\bm H}}
\newcommand{\Ibf}{{\bm I}}
\newcommand{\Jbf}{{\bm J}}

\newcommand{\Lbf}{{\bm L}}

\newcommand{\Obf}{{\bm O}}

\newcommand{\Xbf}{{\bm X}}
\newcommand{\Ybf}{{\bm Y}}
\newcommand{\Zbf}{{\bm Z}}

\newcommand{\abf}{{\bm a}}
\newcommand{\bbf}{{\bm b}}

\newcommand{\xbf}{{\bm x}}

\newcommand{\zbf}{{\bm z}}

\newcommand{\greekbold}[1]{\mbox{\boldmath $#1$}}

\newcommand{\betabf}{\greekbold{\beta}}

\newcommand{\gammabf}{\greekbold{\gamma}}

\newcommand{\thetabf}{\greekbold{\theta}}

\newcommand{\Lambdabf}{\greekbold{\Lambda}}

\newcommand{\Pibf}{\greekbold{\Pi}}

\title{Tensor Regression with Applications \\in Neuroimaging Data Analysis}
\author{Hua Zhou \\
Department of Statistics \\
North Carolina State University \\
Raleigh, NC, 27695-8203 \\
{\tt hua\_zhou@ncsu.edu}
\and
Lexin Li \\
Department of Statistics \\
North Carolina State University \\
Raleigh, NC, 27695-8203 \\
{\tt li@stat.ncsu.edu}
\and
Hongtu Zhu  \\
Department of Biostatistics and \\
Biomedical Research Imaging Center \\
University of North Carolina \\
Chapel Hill, NC 27599-7420 \\
{\tt hzhu@bios.unc.edu}
}
\date{}

\begin{document}
\maketitle

%\begin{footnotetext}[1]
%{Hua Zhou is Assistant Professor, Department of Statistics, North Carolina State University, Raleigh, NC 27695-8203 (Email: hua\_zhou@ncsu.edu).
%Lexin Li is the corresponding author and Associate Professor, Department of Statistics, North Carolina State University, Raleigh, NC 27695-8203 (Email: li@stat.ncsu.edu).
%Hongtu Zhu is Professor, Department of Biostatistics and Biomedical Research Imaging Center, University of North Carolina, Chapel Hill, NC 27599-7420 (E-mail: hzhu@bios.unc.edu).}
%\end{footnotetext}

\begin{abstract}
Classical regression methods treat covariates as a vector and estimate a corresponding vector of regression coefficients. Modern applications in medical imaging generate covariates of more complex form such as multidimensional arrays (tensors). Traditional statistical and computational methods are proving insufficient for analysis of these high-throughput data due to their ultrahigh dimensionality as well as complex structure. In this article, we propose a new family of tensor regression models that efficiently exploit the special structure of tensor covariates. Under this framework, ultrahigh dimensionality is reduced to a manageable level, resulting in efficient estimation and prediction. A fast and highly scalable estimation algorithm is proposed for maximum likelihood estimation and its associated asymptotic properties are studied. Effectiveness of the new methods is demonstrated on both synthetic and real MRI imaging data.
\end{abstract}


\baselineskip=20pt

\section{Introduction}

Understanding the structure and function of the human brains and their connection with neuropsychiatric and neurodegenerative disorders is one of the most intriguing scientific questions \citep{Towle1993,Niedermeyer2004,Buzsaki2006,Fass2008,Lindquist2008,Lazar2008,Friston2009, KangOmbao2012}. Rapidly advancing medical imaging technologies provide powerful tools to help address this question. There are a variety of imaging modalities, including anatomical magnetic resonance imaging (MRI), functional magnetic resonance imaging (fMRI), electroencephalography (EEG), diffusion tensor imaging (DTI), and positron emission tomography (PET), among others. The goal is to understand neural development of both normal brains and brains with mental disorders through one or several imaging modalities. The size and complexity of  medical imaging data, however, are posing unprecedented demands for new statistical methods and theories. In this article, we focus on a family of problems using imaging data to predict cognitive outcome, to classify disease status,  and to identify brain regions associated with clinical response, which have received increasing interest in recent years \citep{Lindquist2008,Lazar2008,  Martino2008, Friston2009, Ryali2010, Hinrichs2009}. The problems can be formulated in a regression setup by treating  clinical outcome as response, and treating images, which are in the form of \emph{multi-dimensional array}, as covariates. However, most classical regression methods take vectors as  covariates. Naively turning an image array into a vector is clearly an unsatisfactory solution.  For instance, with a typical anatomical MRI image of size 256-by-256-by-256, it implicitly requires $256^3 = 16,777,216$ regression parameters. Both computability and theoretical guarantee of the classical regression models are compromised by this ultra-high dimensionality. More seriously, vectorizing an array destroys the inherent spatial structure of the image that possesses wealth of information.

In the literature, there have been roughly three categories of solutions to establishing association between  matrix/array covariates and clinical outcome. The first is the voxel-based methods, which take the image data at each voxel  as responses and clinical variables such as age and gender as predictors, and then generate a statistical parametric map of test statistics or $p$-values across all voxels \citep{Lazar2008, Worsley2004}. A major drawback is that they treat all voxels as independent units, since the fit is at individual voxel level, and thus ignore the fact that voxels are spatially correlated \citep{Li2011, YueLoh2010, Polzehl2010}. The second type of solutions adopts the functional data point of view by taking a one-dimensional function as predictor. Fitting such models commonly involves representing functions as a linear combination of basis functions, which are either pre-specified or obtained from principal component decompositions \citep{RamsaySilverman05FDABook}. \citet{ReissOgden10FunctionalGLM} notably extended this idea to a functional regression model with two-dimensional images as predictors. Extending their method to 3D and higher dimensional images, however, is far from trivial and requires substantial research, due to the large number of parameters and multi-collinearity among imaging measures. The third category employs a two-stage strategy, first carrying out a dimension reduction step, often by principal component analysis (PCA), then fitting a model based on the reduced-dimensional principal components \citep{Caffo2010}. This strategy is intuitive and easy to implement. However, it is well known that PCA is an unsupervised dimension reduction technique and the extracted principal components can be irrelevant to the response. Moreover, theoretical properties of such two-stage solutions are usually intractable and no theoretical results are currently available.

In this article, we propose a new class of regression models for array-valued covariates. Exploiting the array structure in imaging data, the new method substantially reduces the dimensionality of imaging data, which leads to efficient estimation and prediction. The method works for general array-valued covariates and/or any combination of them, and thus it is applicable to a variety of imaging modalities,  e.g., EEG, MRI and fMRI. It is embedded in a generalized linear model (GLM) framework, so it works for both continuous and discrete responses. Within the proposed model framework, we develop both a highly scalable maximum likelihood estimation algorithm as well as statistical inferential tools. Regularized tensor regression is also developed to identify regions of interest in brains that are relevant to a particular response. This \emph{region selection} problem corresponds to \emph{variable selection} in the usual vector-valued regression.

The contributions of this article are two-fold. First, from an image analysis point of view, our proposal timely responds to a number of growing needs of neuroimaging analysis. It also provides a systematic solution for the integrative analysis of multi-modality imaging data and imaging genetics data \citep{Friston2009, Casey2010}. Second, from a statistical methodology point of view, our proposal provides a novel and broad framework for regression with array covariates. A large number of models and extensions are potential outcomes within this framework. Although there has been imaging studies utilizing tensor structure \citep{LiDuLin05, ParkSavvides07}, our proposal, to the best of our knowledge, is the first work that integrates tensor decomposition within a statistical regression (\emph{supervised learning}) paradigm. Our work can be viewed as a logic extension from the classical vector-valued covariate regression to functional covariate regression and then to array-valued covariate regression.

The rest of the article is organized as follows. Section \ref{sec:model} begins with a review of matrix/array properties, and then develops the tensor regression models. Section \ref{sec:estimation} presents an efficient algorithm for maximum likelihood estimation. Section \ref{sec:theory} provides theoretical results such as identifiability, consistency, and asymptotic normality. Section \ref{sec:regularization} discusses regularization including region selection. Section \ref{sec:numerics} presents numerical results. Section \ref{sec:discussion} concludes with a discussion of future extensions. Technical proofs are delegated to the Appendix.

\section{Model}
\label{sec:model}

\subsection{Preliminaries}
\label{sec:notation}

Multidimensional array, also called \emph{tensor}, plays a central role in our approach and we start with a brief summary of  notation and a few results for matrix/array operations. Extensive references can be found in the text \citep{MagnusNeudecker99MatrixBook} for matrix calculus and the survey paper \citep{KoldaBader09Tensor} for tensors. In this article we use the terms multidimensional array and tensor interchangeably.

First we review two matrix products. Given two matrices $\Abf = [\abf_1 \ldots \abf_n] \in \real{m \times n}$ and $\Bbf = [\bbf_1 \ldots \bbf_q] \in \real{p \times q}$, the \emph{Kronecker product} is the $mp$-by-$nq$ matrix
$
    \Abf \otimes \Bbf = [
    \abf_1 \otimes \Bbf \,\, \abf_1 \otimes \Bbf \,\, \ldots \,\, \abf_n \otimes \Bbf ].
$
If $\Abf$ and $\Bbf$ have the same number of columns $n=q$, then the \emph{Khatri-Rao} product \citep{RaoMitra71GenInv} is defined as the $mp$-by-$n$ columnwise Kronecker product
$
    \Abf \odot \Bbf = \left[ \begin{array}{cccc}
        \abf_1 \otimes \bbf_1 & \abf_2 \otimes \bbf_2& \ldots & \abf_n \otimes \bbf_n
    \end{array} \right].
$
If $n=q=1$, then $\Abf \odot \Bbf = \Abf \otimes \Bbf$.
Next, we introduce some useful operations that transform a tensor into a matrix/vector. The \emph{$\vect(\Bbf)$ operator} stacks the entries of a $D$-dimensional tensor $\Bbf \in \real{p_1 \times \cdots \times p_D}$ into a column vector. Specifically, an entry $b_{i_1\ldots i_D}$ maps to the $j$-th entry of $\vect \, \Bbf$, in which
$
    j = 1 + \sum_{d=1}^D (i_d-1) \prod_{d'=1}^{d-1} p_{d'}.
$
For instance, when $D=2$, the matrix entry $x_{i_1i_2}$ maps to position $j= 1 + i_1 -1 + (i_2-1)p_1 = i_1 + (i_2-1)p_1$, which is consistent with the more familiar $\vect$ operation on a matrix. The \emph{mode-$d$ matricization}, $\Bbf_{(d)}$, maps a tensor $\Bbf$ into a $p_d \times \prod_{d' \ne d} p_{d'}$ matrix such that the $(i_1,\ldots,i_D)$ element of the array $\Bbf$ maps to the $(i_d,j)$ element of the matrix $\Bbf_{(d)}$, where
$
    j = 1 + \sum_{d'\ne d} (i_{d'}-1) \prod_{d''<d',d'' \ne d} p_{d''}.
$
With $d=1$, we observe that $\vect \, \Bbf$ is the  same as vectorizing the mode-1 matricization $\Bbf_{(1)}$. The \emph{mode-($d,d'$) matricization} $\Bbf_{(dd')} \in \real{p_dp_{d'} \times \prod_{d'' \ne d,d'} p_{d''}}$ is defined in a similar fashion \citep{Kolda06multilinearoperators}.
We also introduce an operator that turns vectors into an array. Specifically, an \emph{outer product}, $\bbf_1 \circ \bbf_2 \circ \cdots \circ \bbf_D$, of $D$ vectors $\bbf_d \in \real{p_d}$, $d=1,\ldots,D$, is a $p_1 \times \cdots \times p_D$ array with entries $(\bbf_1 \circ \bbf_2 \circ \cdots \circ \bbf_D)_{i_1 \cdots i_D} = \prod_{d=1}^D b_{di_d}$.

Next we introduce a concept that plays a central role in our proposed tensor regression in Section \ref{sec:tensor-model}. We say an array $\Bbf \in \real{p_1 \times \cdots \times p_D}$ admits a \emph{rank-$R$ decomposition} if
\begin{align}
    \Bbf = \sum_{r=1}^R \betabf_1^{(r)} \circ \cdots \circ \betabf_D^{(r)},   \label{eqn:R-CP-decomp}
\end{align}
where $\betabf_d^{(r)} \in \real{p_d}, d=1,\ldots,D,r=1,\ldots,R$, are all column vectors, and $\Bbf$ cannot be written as a sum of less than $R$ outer products. For convenience, the decomposition is often represented by a shorthand, $\Bbf = \llbracket \Bbf_1,\ldots,\Bbf_D \rrbracket$,  where $\Bbf_d = [\betabf_d^{(1)}, \ldots, \betabf_d^{(R)}] \in \real{p_d \times R}$, $d=1, \ldots, D$ \citep{Kolda06multilinearoperators,KoldaBader09Tensor}. The following well-known result relates the mode-$d$ matricization and the vec operator of an array to its rank-$R$ decomposition. The proof is given in the Appendix for completeness.
\begin{lemma}
\label{lemma:mode-d-matricize}
If a tensor $\Bbf \in \real{p_1 \times \cdots \times p_D}$ admits a rank-$R$ decomposition (\ref{eqn:R-CP-decomp}), then
\begin{align*}
    \Bbf_{(d)} &= \Bbf_d (\Bbf_D \odot \cdots \odot \Bbf_{d+1} \odot \Bbf_{d-1} \odot \cdots \odot \Bbf_1) \trans  \, \text{ and}~~
    \vect \, \Bbf    =   (\Bbf_D \odot \cdots \odot \Bbf_1) {\bf 1}_{R}.
\end{align*}
\end{lemma}

Throughout the article, we adopt the following notations. $Y$ is a univariate response variable, $\Zbf \in \real{p_0}$ denotes a $p_0$-dimensional   vector of covariates, such as   age and  sex, and  $\Xbf \in \real{p_1 \times \ldots \times p_D}$ is a $D$-dimensional array-valued predictor. For instance, for MRI, $D=3$, representing the 3D structure of an image, whereas  for fMRI, $D = 4$, with an additional time dimension. The lower-case triplets $(y_i, \xbf_i, \zbf_i)$, $i=1,\ldots,n$, denote the independent, observed sample instances of $(Y, \Xbf, \Zbf)$.

\subsection{Motivation and Basic Model}
\label{sec:basic-model}

To motivate our model, we first start with a vector-valued $\Xbf$ and absorb $\Zbf$ into $\Xbf$. In the classical GLM \citep{McCullaghNelder83GLMBook} setting, $Y$ belongs to an exponential family with probability mass function or density,
\begin{eqnarray} \label{GLM}
p(y|\theta,\phi) = \exp\left\{ \frac{y \theta - b(\theta)}{a(\phi)} + c(y,\phi) \right\} \label{eqn:GLM-density}
\end{eqnarray}
where $\theta$  and $\phi>0$ denote the natural and dispersion parameters. The classical GLM relates a vector-valued $\Xbf \in \real{p}$ to the mean $\mu = E(Y|\Xbf)$ via  $g(\mu) = \eta =  \alpha + \betabf\trans \Xbf$, where $g(\cdot)$ is a strictly increasing link function, and $\eta$ denotes the linear systematic part with intercept $\alpha$ and the coefficient vector $\betabf \in \real{p}$.

Next, for a matrix-valued covariate $\Xbf \in \real{p_1 \times p_2}$ ($D = 2$), it is intuitive to consider a GLM model with the systematic part given by
\begin{eqnarray*}
\label{eqn:motivation}
    g(\mu) =  \alpha  + \betabf_1 \trans \Xbf \betabf_2,
\end{eqnarray*}
where $\betabf_1 \in \real{p_1}$ and $\betabf_2 \in \real{p_2}$, respectively.  The bilinear form $\betabf_1 \trans \Xbf \betabf_2$ is a natural extension of the linear term $\beta\trans \Xbf$ in the classical GLM with a vector covariate $\Xbf$. It is interesting to note that, this bilinear form was first proposed by \citet{LiKimAltman2010} in the context of dimension reduction, and then employed by \citet{HungWang11} in the logistic regression with matrix-valued covariates ($D=2$). Moreover, note that $\betabf_1\trans \Xbf \betabf_2 = (\betabf_2 \otimes \betabf_1)\trans \vect(\Xbf)$.

Now for a conventional vector-valued covariate $\Zbf$ and a general array-valued $\Xbf \in \real{p_1 \times \ldots \times p_D}$, we   propose a GLM  with the systematic part given by
\begin{eqnarray}
\label{eqn:model}
g(\mu) = \alpha +  \gammabf \trans \Zbf + (\betabf_D \otimes \ldots \otimes \betabf_1) \trans \vect(\Xbf),
\end{eqnarray}
where $\gammabf \in \real{p_0}$ and $\betabf_d \in \real{p_d}$ for $d = 1, \ldots, D$. This is our \emph{basic model} for regression with array covariates. The key advantage of  model  (\ref{eqn:model}) is that it dramatically and effectively reduces the dimensionality of the tensor component, from the order of $\prod_d p_d$ to the order of $\sum_d p_d$. Take MRI imaging as an example, the size of a typical image is $256^3 =  16,777,216$. If we simply turn $\Xbf$ into a vector and fit a GLM, this brutal force solution is over 16 million-dimensional, and the computation is practically infeasible. By contrast,  (\ref{eqn:model}) turns the problem to  be $256+256+256 = 768$-dimensional. The reduction in dimension, and consequently in computational saving, is substantial.

\begin{figure}
\begin{center}
\begin{tabular}{c}
\includegraphics[width=4.4in]{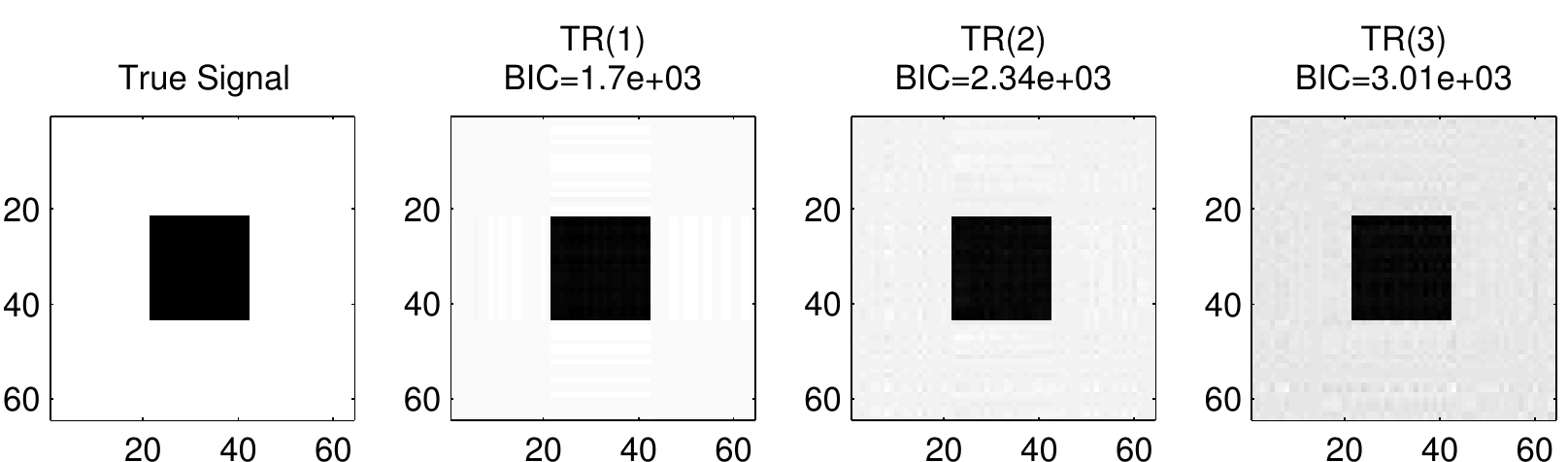} \\
\includegraphics[width=4.4in]{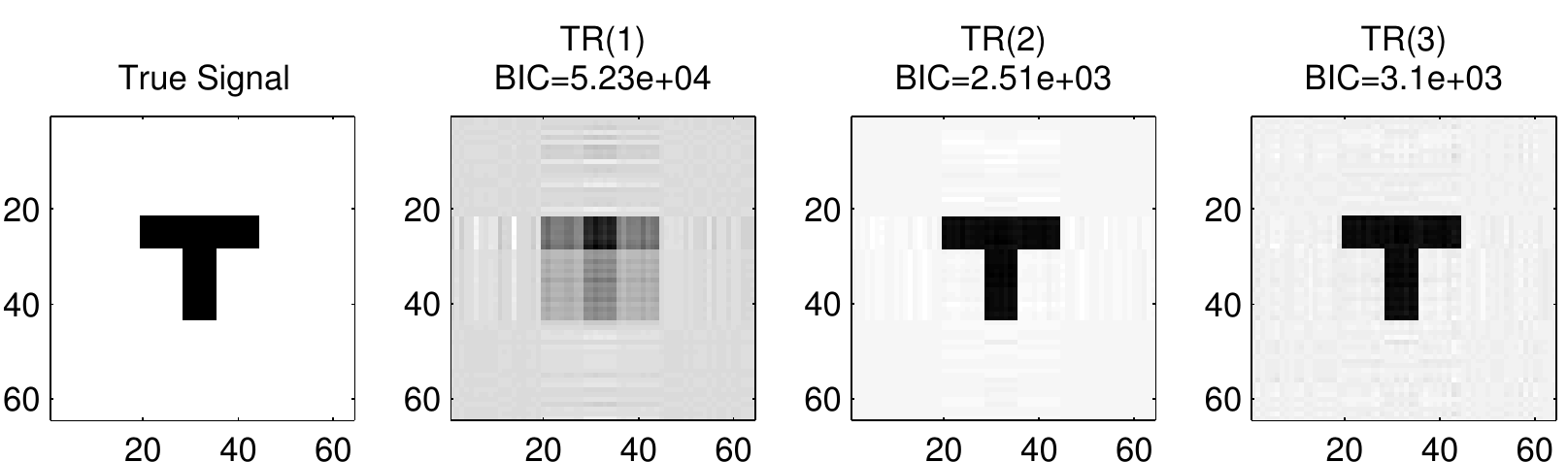} \\
\includegraphics[width=4.4in]{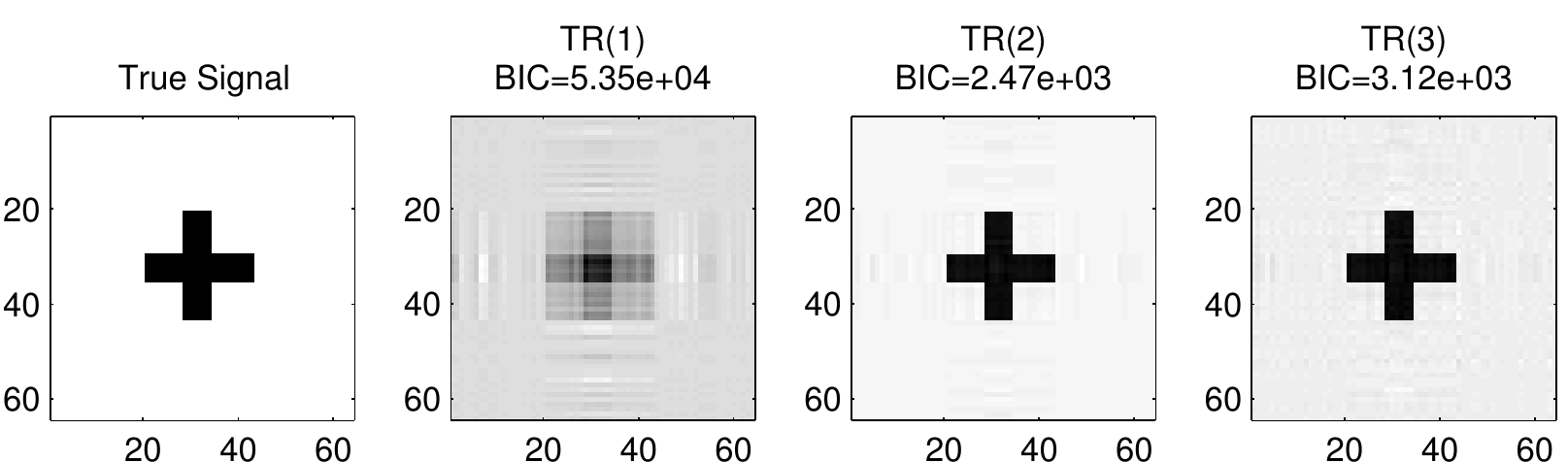} \\
\includegraphics[width=4.4in]{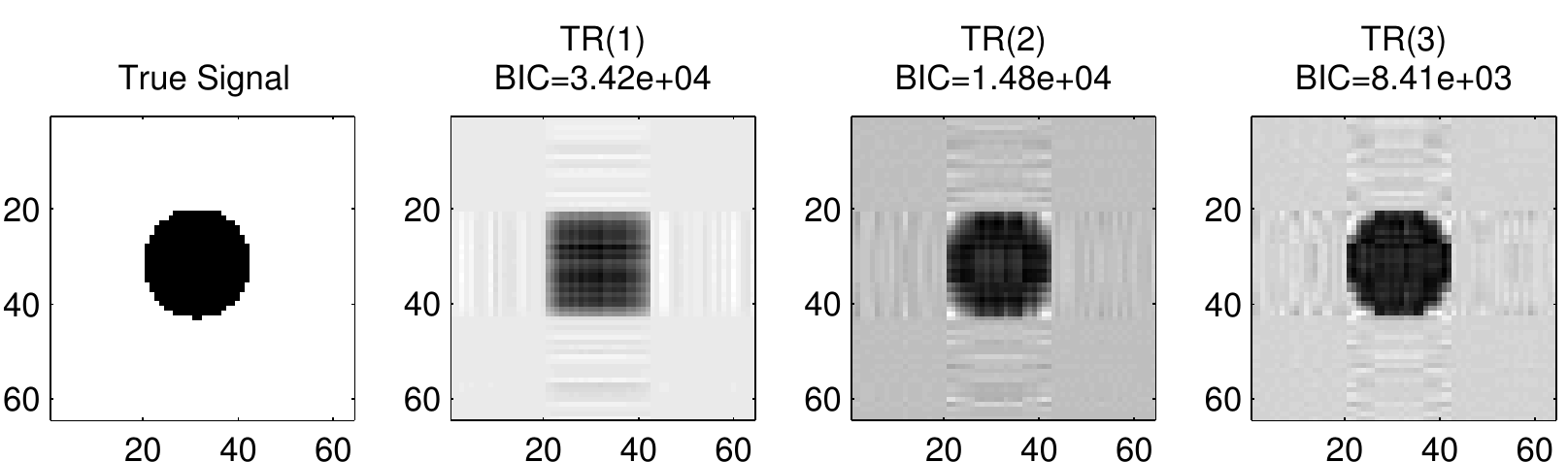} \\
\includegraphics[width=4.4in]{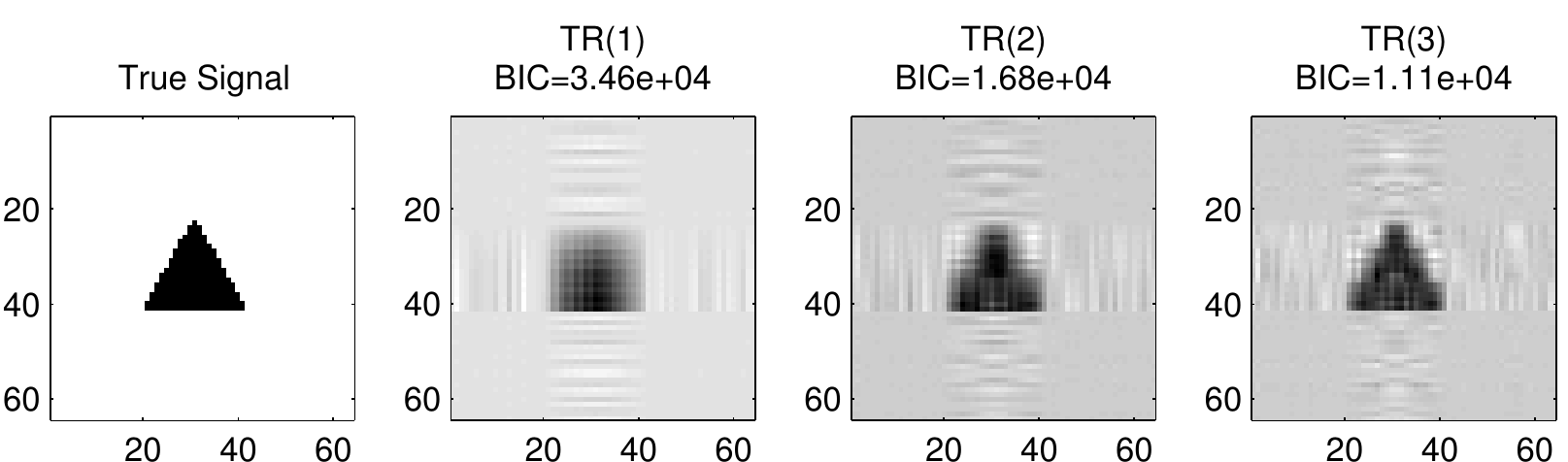} \\
\includegraphics[width=4.4in]{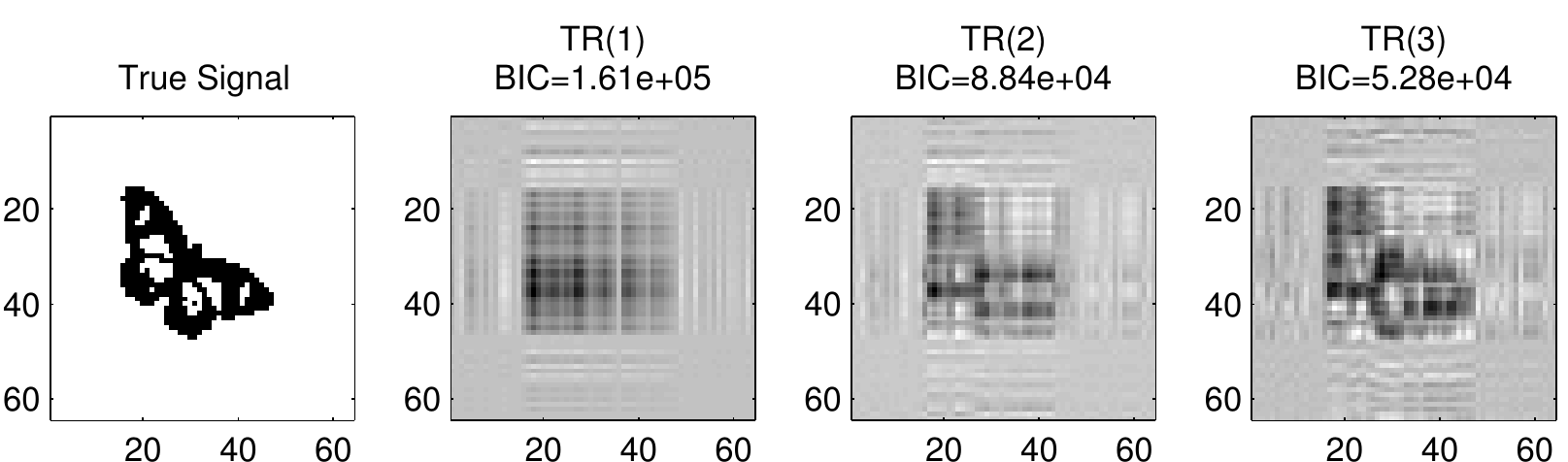}
\end{tabular}
\caption{True and recovered image signals by tensor regression. The matrix variate has size 64 by 64 with entries generated as independent standard normals. The regression coefficient for each entry is either 0 (white) or 1 (black). The sample size is 1000. TR$(R)$ means estimate from the rank-$R$ tensor regression.}\label{fig:shapes}
\end{center}
\end{figure}

A critical question then is whether such a massive reduction in the number of parameters would limit the capacity of model  (\ref{eqn:model})  to capture  regions of interest with specific shapes. The illustrative example in Figure \ref{fig:shapes} provides some clues.  In Figure \ref{fig:shapes}, we present  several two-dimensional images $\Bbf \in \real{64 \times 64}$ (shown in the first column), along with the estimated images by model (\ref{eqn:model}) (in the second column labeled by TR(1)). Specifically, we simulated 1,000 univariate responses $y_i$ according to  a normal model with mean $\mu_i = \gammabf \trans \zbf_i + \langle \Bbf, \xbf_i \rangle$, where $\gammabf = {\bf 1}_5$. The inner product between two arrays is defined as $\langle \Bbf,\Xbf \rangle = \langle \vect \Bbf, \vect \Xbf \rangle = \sum_{i_1,\ldots,i_D} \beta_{i_1\ldots i_D} x_{i_1 \ldots i_D}$. The coefficient array $\Bbf$ is binary, with the true signal region equal to one and the rest zero. The regular covariate $\zbf_i$ and image covariate $\xbf_i$ are randomly generated with all elements being independent standard normals. Our goal is to see if model (\ref{eqn:model}) can identify the true signal region in $\Bbf$ using data $(y_i,\zbf_i,\xbf_i)$. Before examining the outcome, we make two remarks about this illustration. First, our problem differs from the usual edge detection or object recognition in imaging processing \citep{Qiu2005book,Qiu07jumpsurface}. In our setup, all elements of the image $\Xbf$ follow the same distribution. The signal region is defined through the coefficient image $\Bbf$ and needs to be inferred from the association between $Y$ and $\Xbf$ after adjusting for $\Zbf$. Second, the classical GLM is difficult to apply in this example if we simply treat $\mbox{vec}(\Xbf)$ as a covariate vector, since the sample size $n=1,000$ is much less than the number of parameters $p=5 + 64 \times 64 = 4,101$. Back to Figure \ref{fig:shapes}, the second column clearly demonstrates the ability of model (\ref{eqn:model}) in identifying the rectangular (square) type region (parallel to the image edges). On the other hand, since the parameter vector $\betabf_d$ in a rank-1 model is only able to capture the accumulative signal along the $d$-th dimension of the array variate $\Xbf$, it is unsurprising that it does not perform well for signals that are far away from rectangle, such as triangle, disk, T-shape and butterfly. This motivates us to develop a more flexible tensor regression model in the next section.

\subsection{Tensor Regression Model}
\label{sec:tensor-model}

We start with an alternative view of the basic model (\ref{eqn:model}), which will lead to its generalization. Consider a $D$-dimensional array variate $\Xbf \in \real{p_1 \times \cdots \times p_D}$, and a full coefficient array $\Bbf$ of same size that captures the effects of each array element. Then the most flexible GLM suggests a linear systematic part
\begin{eqnarray*}
    g(\mu) =  \alpha + \gamma \trans \Zbf + \langle \Bbf,\Xbf \rangle.
\end{eqnarray*}
The issue with this model is that $\Bbf$ has the same number of parameters, $\prod_{d=1}^D p_d$, as $\Xbf$, which is ultrahigh dimensional and far exceeds the usual sample size. Then a natural idea is to approximate $\Bbf$ with less parameters. If $\Bbf$ admits a rank-1 decomposition (\ref{eqn:R-CP-decomp}), i.e., $\Bbf = \betabf_1 \circ \betabf_2 \circ \cdots \circ \betabf_D$, where $\beta_d \in \real{p_d}$, then by Lemma \ref{lemma:mode-d-matricize}, we have
\begin{eqnarray*}
\vect \, \Bbf = \vect \left(\betabf_1 \circ \betabf_2 \circ \cdots \circ \betabf_D\right) = \betabf_D \odot \cdots \odot \betabf_1 = \betabf_D \otimes \cdots \otimes \betabf_1.
\end{eqnarray*}
In other words, model (\ref{eqn:model}) is indeed a \emph{data-driven} model with a rank-1 approximation to the general signal array $\Bbf$. This observation motivates us to consider a more flexible   tensor regression model.

Specifically, we propose a family of \emph{rank-$R$ generalized linear tensor regression models}, in which the systematic part of GLM is of the form
\begin{align}
    g(\mu) &=  \alpha + \gamma \trans \Zbf +  \langle \sum_{r=1}^R \betabf_1^{(r)} \circ \betabf_2^{(r)} \circ \cdots \circ \betabf_D^{(r)},\Xbf  \rangle  \nonumber \\
   &= \alpha + \gamma \trans \Zbf + \langle (\Bbf_D \odot \cdots \odot \Bbf_1) {\bf 1}_{R} , \vect \Xbf \rangle, \label{eqn:r-tensorreg}
\end{align}
where $\Bbf =   \llbracket \Bbf_1,\ldots,\Bbf_D \rrbracket = \sum_{r=1}^R \betabf_1^{(r)} \circ \betabf_2^{(r)} \circ \cdots \circ \betabf_D^{(r)}$; i.e., it admits a rank-$R$ decomposition, $\Bbf_d = [\betabf_d^{(1)}, \ldots, \betabf_d^{(R)}] \in \real{p_d \times R}$, $\Bbf_D \odot \cdots \odot \Bbf_1 \in \real{\prod_d p_d \times R}$ is the Khatri-Rao product and ${\bf 1}_{R}$ is the vector of $R$ ones.  When $R=1$, it reduces to   model (\ref{eqn:model}). A few remarks on \eqref{eqn:r-tensorreg} are in order. First, since our formulation only deals with the linear predictor part of the model, it easily extends to the quasi-likelihood models \citep{McCullaghNelder83GLMBook} where more general mean-variance relation is assumed. Second, for simplicity, we only discuss exponential family with a univariate response. Extension to multivariate exponential family, such as multinomial logit model, is straightforward. Third, due to the GLM setup (\ref{eqn:GLM-density}), we call \eqref{eqn:r-tensorreg} a generalized linear tensor regression model. However, we should bear in mind that the systematic component $\eta$ is a polynomial rather than linear in the parameters $\Bbf_d$. Finally, the rank-$R$ tensor decomposition  (\ref{eqn:R-CP-decomp}) is called canonical decomposition or parallel factors (CANDECOMP/PARAFAC, or CP) in psychometrics \citep{KoldaBader09Tensor}. In that sense, model (\ref{eqn:r-tensorreg}) can be viewed as a \emph{supervised} version of the classical CP decomposition for multi-dimensional arrays.

The number of parameters in model (\ref{eqn:r-tensorreg}) is $p_0 + R \sum_d p_d$, which is still substantially smaller than $p_0 + \prod_d p_d$. With such a massive reduction in dimensionality, however, we demonstrate it provides a reasonable approximation to many low rank signals. Returning to the previous illustration, in Figure \ref{fig:shapes}, images TR($R$) are the recovered signals by the rank-$R$ tensor regression (in third and fourth columns). The square signal can be perfectly recovered by a rank-1 model, whereas rank-2 and 3 regressions show signs of overfitting. The T-shape and cross signals can be perfectly recovered by a rank-2 regression. Triangle, disk, and butterfly shapes cannot be exactly recovered by any low rank approximations; however, a rank 3 tensor regression already yields a fairly accurate recovery. Clearly, the general tensor regression model (\ref{eqn:r-tensorreg}) is able to capture significantly more tensor signals than the basic model (\ref{eqn:model}).

\section{Estimation}
\label{sec:estimation}

We pursue the maximum likelihood (ML) route for parameter estimation in model (\ref{eqn:r-tensorreg}).   Given  $n$ i.i.d.\ data $\{(y_i, \xbf_i, \zbf_i), i=1,\ldots,n\}$, the log-likelihood function for   (\ref{eqn:GLM-density}) is
\begin{eqnarray}
\ell(\alpha,\gammabf,\Bbf_1,\ldots,\Bbf_D) = \sum_{i=1}^n \frac{y_i \theta_i - b(\theta_i)}{a(\phi)} + \sum_{i=1}^n c(y_i,\phi), \label{eqn:glm-loss}
\end{eqnarray}
where $\theta_i$ is related to regression parameters $(\alpha,\gammabf,\Bbf_1,\ldots,\Bbf_D)$ through (\ref{eqn:r-tensorreg}).
We propose an efficient algorithm for maximizing $\ell(\alpha,\gammabf,\Bbf_1,\ldots,\Bbf_D)$. A key observation is that although $g(\mu)$ in (\ref{eqn:r-tensorreg}) is not linear in $(\Bbf_1, \ldots,\Bbf_D)$ jointly, it is linear in $\Bbf_d$ individually. This suggests alternately updating $(\alpha,\gammabf)$ and $\Bbf_d$, $d=1,\ldots,D$, while keeping other components fixed. It yields a so-called \emph{block relaxation algorithm} \citep{deLeeuw94BR,Lange10NumAnalBook}. An appealing feature of this algorithm is that at each iteration, updating a block $\Bbf_d$ is simply a classical GLM problem. To see this, when updating $\Bbf_d \in \real{p_d \times R}$, we rewrite the array inner product in (\ref{eqn:r-tensorreg}) as
\begin{eqnarray*}
      \langle \sum_{r=1}^R \betabf_1^{(r)} \circ \betabf_2^{(r)} \circ \cdots \circ \betabf_D^{(r)},\Xbf \rangle
    = \langle \Bbf_d, \Xbf_{(d)} (\Bbf_D \odot \cdots \odot \Bbf_{d+1} \odot \Bbf_{d-1} \odot \cdots \odot \Bbf_1) \rangle.
\end{eqnarray*}
Consequently the problem turns into a traditional GLM regression with $Rp_d$ parameters, and the estimation procedure breaks into a sequence of low dimensional GLM optimizations and is extremely easy to implement using ready statistical softwares such as {\sc R}, {\sc S+}, {\sc SAS}, and {\sc Matlab}. The full estimation procedure is summarized in Algorithm \ref{algo:br-algo}. For the Gaussian models, it reduces to the alternating least squares (ALS) procedure \citep{deLeeuwYoungTakane76ALS}.

\begin{algorithm}[t]
\begin{algorithmic}
\STATE Initialize: $(\alpha^{(0)},\gammabf^{(0)}) = \mbox{argmax}_{\alpha,\gammabf} \, \ell(\alpha,\gammabf,{\bf 0}, \ldots, {\bf 0})$, $\Bbf_d^{(0)} \in $ $\real{p_d \times R}$ a random matrix for $d=1,\ldots,D$.
\REPEAT
\FOR{$d=1, \ldots, D$}
\STATE $\Bbf_d^{(t+1)} = \mbox{argmax}_{\Bbf_d} \, \ell(\alpha^{(t)},\gammabf^{(t)}, \Bbf_1^{(t+1)}, \ldots, \Bbf_{d-1}^{(t+1)}, \Bbf_d, \Bbf_{d+1}^{(t)}, \ldots, \Bbf_D^{(t)})$
\ENDFOR
\STATE $(\alpha^{(t+1)},\gammabf^{(t+1)}) = \mbox{argmax}_{\alpha,\gammabf} \, \ell(\alpha,\gammabf, \Bbf_1^{(t+1)}, \ldots, \Bbf_D^{(t+1)})$
\UNTIL{$\ell(\thetabf^{(t+1)})-\ell(\thetabf^{(t)}) < \epsilon$}
\end{algorithmic}
\caption{Block relaxation algorithm for maximizing (\ref{eqn:glm-loss}).}
\label{algo:br-algo}
\end{algorithm}

As the block relaxation algorithm monotonically increases the objective function, it is numerically stable and the convergence of objective values $\ell(\thetabf^{(t)})$ is guaranteed whenever $\ell(\thetabf)$ is bounded from above. Therefore the stopping rule of Algorithm \ref{algo:br-algo} is well-defined. We denote the algorithmic map by $M$, i.e., $M(\thetabf^{(t)}) = \thetabf^{(t+1)}$, with $\thetabf=(\alpha,\gammabf,\Bbf_1,\ldots,\Bbf_D)$ collecting all parameters.  Convergence properties of Algorithm \ref{algo:br-algo} are summarized in Proposition \ref{prop:glob-conv}. The proof is relegated to the Appendix.
\begin{proposition}
\label{prop:glob-conv}
Assume (i) the log-likelihood function $\ell(\thetabf)$ is continuous, coercive, i.e., the set $\{\thetabf: \ell(\thetabf) \ge \ell(\thetabf^{(0)})\}$ is compact, and bounded above, (ii) the objective function in each block update of Algorithm \ref{algo:br-algo} is strictly concave, and (iii) the set of stationary points (modulo scaling and permutation indeterminancy) of $\ell(\thetabf)$ are isolated. We have the following results.
\begin{enumerate}
\item (Global Convergence) The sequence $\thetabf^{(t)} = (\alpha^{(t)}, \gammabf^{(t)}, \Bbf_1^{(t)}, \ldots, \Bbf_D^{(t)})$ generated by Algorithm \ref{algo:br-algo} converges to a stationary point of $\ell(\thetabf)$.
\item (Local Convergence) Let $\thetabf^{(\infty)} = (\alpha^{(\infty)},\gammabf^{(\infty)},\Bbf_1^{(\infty)},\ldots,\Bbf_D^{(\infty)})$ be a strict local maximum of $\ell(\thetabf)$. The iterates generated by Algorithm \ref{algo:br-algo} are locally attracted to $\thetabf^{(\infty)}$ for $\thetabf^{(0)}$ sufficiently close to $\thetabf^{(\infty)}$.
\end{enumerate}
\end{proposition}
We make a few quick remarks. First, although a stationary point is not guaranteed to be even a local maximum (it can be a saddle point), in practice the block relaxation algorithm almost always converges to at least a local maximum. In general, the algorithm should be run from multiple initializations to locate an excellent local maximum. Second, $\ell(\thetabf)$ is not required to be jointly concave in $\thetabf$, but only the concavity in the blocks of variables is needed. This condition holds for all GLM with canonical link such as linear regression and Poisson regression with exponential link.

The above algorithm assumes a known rank when estimating $\Bbf$. Estimating an appropriate rank for our tensor model (\ref{eqn:r-tensorreg}) is of practical importance. It can be formulated as a model selection problem, and we adopt the usual model section criterion, e.g., Bayesian information criterion (BIC), $-2 \ell(\thetabf) + \log(n) p_e$, where $p_e$ is the effective number of parameters for model (\ref{eqn:r-tensorreg}): $p_e = R(p_1+p_2) - R^2$ for $D=2$, and $p_e = R(\sum_d p_d - D + 1)$ for $D > 2$. Returning to the illustrative example in Section \ref{sec:basic-model}, we fitted a rank-1, 2 and 3 tensor models, respectively, to various signal shapes. The corresponding BIC values are shown in Figure \ref{fig:shapes}. The criterion is seen correctly estimating the rank for square as 1, and the rank for T and cross as 2. The true ranks for disk, triangle and butterfly are above 3, and their BIC values at rank 3 are smallest compared to those at 1 and 2.

\section{Theory}
\label{sec:theory}

We study the statistical properties of maximum likelihood estimate (MLE) for the tensor regression model defined by (\ref{eqn:GLM-density}) and (\ref{eqn:r-tensorreg}). For simplicity, we omit the intercept $\alpha$ and the classical covariate part $\gammabf \trans \Zbf$, though the conclusions generalize to an arbitrary combination of covariates. We adopt the usual asymptotic setup with a fixed number of parameters $p$ and a diverging sample size $n$, because this is an important first step toward a comprehensive understanding of the theoretical properties of the proposed model. The asymptotics with a diverging $p$ is our future work and is pursued elsewhere.

\subsection{Score and Information}

We first derive the score and information for the tensor regression model, which are essential for statistical estimation and inference. The following standard calculus notations are used. For a scalar function $f$, $\nabla f$ is the (column) gradient vector, $df = [\nabla f] \trans$ is the differential, and $d^2f$ is the Hessian matrix. For a multivariate function $g: \real{p} \mapsto \real{q}$, $Dg \in \real{q \times p}$ denotes the Jacobian matrix holding partial derivatives $\partial g_i / \partial x_j$.

We start from the Jacobian and Hessian of the systematic part $\eta \equiv g(\mu)$ in (\ref{eqn:r-tensorreg}). The proof is given in the Appendix.
\begin{lemma}
\label{prop:eta-derivatives}
\begin{enumerate}
\item The gradient $\nabla \eta(\Bbf_1,\ldots,\Bbf_D) \in \real{R \sum_{d=1}^D p_d}$ is
\begin{eqnarray*}
    \nabla \eta(\Bbf_1,\ldots,\Bbf_D) = [\Jbf_1 \,\, \Jbf_2 \,\, \cdots \,\, \Jbf_D] \trans (\vect \Xbf),
\end{eqnarray*}
where $\Jbf_d \in \real{\prod_{d=1}^D p_d \times p_dR}$ is the Jacobian
\begin{eqnarray}
    \Jbf_d = D\Bbf(\Bbf_d) =  \Pibf_d [(\Bbf_D \odot \cdots \odot \Bbf_{d+1} \odot \Bbf_{d-1} \odot \cdots \odot \Bbf_1) \otimes \Ibf_{p_d}] \label{eqn:Jd}
\end{eqnarray}
and $\Pibf_d$ is the $(\prod_{d=1}^D p_d)$-by-$(\prod_{d=1}^D p_d)$ permutation matrix that reorders $\vect \Bbf_{(d)}$ to obtain $\vect \Bbf$, i.e., $
    \vect \Bbf = \Pibf_d \, \vect \Bbf_{(d)}.
$
\item The Hessian $d^2 \eta(\Bbf_1,\ldots,\Bbf_D) \in \real{R \sum_{d=1}^D p_d \times R \sum_{d=1}^D p_d}$ has entries
\begin{align*}
    h_{(i_d,r),(i_{d'},r')} &= 1_{\{r=r',d\ne d'\}} \sum_{j_d=i_d,j_{d'}=i_{d'}} x_{j_1,\ldots,j_D} \prod_{d''\ne d, d'} \beta_{j_{d''}}^{(r)},
\end{align*}
and can be partitioned in $D^2$ blocks as
\begin{eqnarray*}
    \left( \begin{array}{cccc}
    {\bf 0} & * & * & * \\
    \Hbf_{21} & {\bf 0} & * & *  \\
    \vdots & \vdots & \ddots & *  \\
    \Hbf_{D1} & \Hbf_{D2} & \cdots & {\bf 0}
    \end{array}
    \right).
\end{eqnarray*}
The block $\Hbf_{dd'} \in \real{p_dR \times p_{d'}R}$ has $p_dp_{d'}R$ nonzero elements which can be retrieved from the matrix
$
    \Xbf_{(dd')} (\Bbf_D \odot \cdots \odot \Bbf_{d+1} \odot \Bbf_{d-1} \odot \cdots \odot \Bbf_{d'+1} \odot \Bbf_{d'-1} \odot \cdots \odot \Bbf_1),
$
where $\Xbf_{(dd')}$ is the mode-$(d,d')$ matricization of $\Xbf$.
\end{enumerate}
\end{lemma}
\noindent
\textit{Remark 1: }
The Hessian $d^2 \eta$ is highly sparse and structured. An entry in $d^2 \eta(\Bbf_1,\ldots,\Bbf_D)$ is nonzero only if it belongs to different directions $d$ but the same outer product $r$.

Let
$
    \ell(\Bbf_1,\ldots,\Bbf_D|y,\xbf) = \ln p (y|\xbf,\Bbf_1,\ldots,\Bbf_D)
$
be the log-density. Next result derives the score function, Hessian, and Fisher information of the tensor regression model.
\begin{proposition}
\label{prop:score-info}
Consider the tensor regression model defined by (\ref{eqn:GLM-density}) and (\ref{eqn:r-tensorreg}).
\begin{enumerate}
\item The score function (or score vector) is
\begin{align}
    \nabla \ell(\Bbf_1,\ldots,\Bbf_D) = \frac{(y - \mu) \mu'(\eta)}{\sigma^2} [\Jbf_1 \ldots \Jbf_D] \trans (\vect \Xbf)    \label{eqn:tensor-score}
\end{align}
with $\Jbf_d$, $d=1,\ldots,D$, defined by (\ref{eqn:Jd}).
\item The Hessian of the log-density $\ell$ is
\begin{align}
    H(\Bbf_1,\ldots,\Bbf_D) ={}& - \frac{[\mu'(\eta)]^2}{\sigma^2} ([\Jbf_1 \ldots \Jbf_D] \trans\vect \Xbf) ([\Jbf_1 \ldots \Jbf_D] \trans\vect \Xbf) \trans   \nonumber \\
    & + \frac{(y-\mu)\theta''(\eta)}{\sigma^2} ([\Jbf_1 \ldots \Jbf_D] \trans\vect \Xbf) ([\Jbf_1 \ldots \Jbf_D] \trans\vect \Xbf) \trans  \nonumber \\
    & + \frac{(y-\mu)\theta'(\eta)}{\sigma^2} d^2 \eta(\Bbf_1,\ldots,\Bbf_D), \label{eqn:tensor-hessian}
\end{align}
with $d^2\eta$ defined in Lemma \ref{prop:eta-derivatives}.
\item The Fisher information matrix is
\begin{eqnarray}
\Ibf(\Bbf_1,\ldots,\Bbf_D) \nonumber & = & E[- H(\Bbf_1,\ldots,\Bbf_D)] = \mathrm{Var} [ \nabla \ell(\Bbf_1,\ldots,\Bbf_D) d\ell(\Bbf_1,\ldots,\Bbf_D)] \nonumber \\
    & = & \frac{[\mu'(\eta)]^2}{\sigma^2}  [\Jbf_1 \ldots \Jbf_D] \trans (\vect \Xbf) (\vect \Xbf) \trans [\Jbf_1 \ldots \Jbf_D]. \label{eqn:fisher-info}
\end{eqnarray}
\end{enumerate}
\end{proposition}

\noindent
\textit{Remark 2: }
For canonical link, $\theta=\eta$, $\theta'(\eta)=1$, $\theta''(\eta)=0$, and the second term of Hessian vanishes. For the classical GLM with linear systematic part ($D=1$), $d^2 \eta(\Bbf_1,\ldots,\Bbf_D)$ is zero and  thus the third term of Hessian vanishes. For the classical GLM ($D=1$) with canonical link, both the second and third terms of the Hessian vanish and thus the Hessian is non-stochastic, coinciding with the information matrix.

\subsection{Identifiability}
\label{sec:identifiability}

Before studying asymptotic property, we need to deal with the identifiability issue. The parameterization in the tensor model is nonidentifiable due to two complications. Consider a rank-$R$ decomposition of an array, $\Bbf = \llbracket \Bbf_1, \ldots, \Bbf_D \rrbracket$. The first complication is the indeterminacy of $   \Bbf$  due to scaling and permutation:
\begin{itemize}
\item[--] scaling: $\Bbf = \llbracket \Bbf_1 \Lambdabf_1, \ldots, \Bbf_D \Lambdabf_D \rrbracket$ for any diagonal matrices $\Lambdabf_d=\text{diag}(\lambda_{d1},\ldots,\lambda_{dR})$, $d=1,\ldots,D$, such that $\prod_{d} \lambda_{dr}=1$ for $r=1,\ldots,R$.
\item[--] permutation: $\Bbf = \llbracket \Bbf_1 \Pibf, \ldots, \Bbf_D \Pibf \rrbracket$ for any $R$-by-$R$ permutation matrix $\Pibf$.
\end{itemize}
For the matrix case ($D=2$), a further complication is the nonsingular transformation indeterminancy: $\Bbf_1 \Bbf_2 \trans = \Bbf_1 \Obf \Obf^{-1} \Bbf_2 \trans$ for any $R$-by-$R$ nonsingular matrix $\Obf$. Note the scaling and permutation indeterminancy is subsumed in the nonsingular transformation indeterminancy. The singular value decomposition (SVD) of a matrix is unique because it imposes orthonormality constraint on the columns of the factor matrices.

To deal with this complication, it is necessary to adopt a specific constrained parameterization to fix the scaling and permutation indeterminacy. For $D>2$, we need to put $(D-1)R$ restrictions on the parameters $\Bbf$ and apparently there is an infinite number of ways to do this. In this paper we adopt the following convention. $\Bbf_1, \ldots, \Bbf_{D-1}$ are scaled such that $\beta_{d1}^{(r)} = 1$, i.e., the first rows are ones. This in turn determines entries in the first row of $\Bbf_D$ and fixes scaling indeterminacy. To fix the permutation indeterminancy, we assume that the first row entries of $\Bbf_D$ are distinct and arranged in descending order $\beta_{D1}^{(1)} > \cdots > \beta_{D1}^{(R)}$. The resulting parameter space is
\begin{align*}
    \mathcal{\Bbf} = \{(\Bbf_1,\ldots,\Bbf_D): & \, \beta_{d1}^{(r)}=1, \text{ for } d=1,\ldots,D, r=1,\ldots,R \text{ and }\beta_{D1}^{(1)} > \cdots > \beta_{D1}^{(R)}\},
\end{align*}
which is open and convex. The formulae for score, Hessian and information in Proposition \ref{prop:score-info} require changes accordingly, i.e., the entries in the first rows of $\Bbf_d$, $d=1,\ldots,D-1$, are fixed at ones and their corresponding entries, rows and columns in score, Hessian and information need to be deleted. Treatment for the $D=2$ case is similar and omitted for brevity.
We emphasize that our choice of the restricted space $\mathcal{\Bbf}$ is arbitrary and exclude many arrays that might be of interest, e.g., arrays with any entries in the first rows of $\Bbf_d$, $d=1,\ldots,D-1$, equal to zeros or with ties in the first row of $\Bbf_D$. However the set of such exceptional arrays has Lebesgue measure zero. In specific applications, subject knowledge may suggest alternative restrictions on the parameters.

The second complication comes from possible non-uniqueness of decomposition when $D>2$ even after adjusting scaling and permutation indeterminacy. The next proposition collects some recent results that give easy-to-check conditions for the uniqueness (up to scaling and permutation) of decomposition. The first two are useful for checking uniqueness of a given tensor, while the latter two give general conditions for uniqueness almost everywhere in the $D=3$ or 4 case.
\begin{proposition}
\label{prop:CP-uniqueness}
Suppose that a $D$-dimensional array $\Bbf \in \real{p_1 \times \cdots \times p_D}$ has rank $R$.
\begin{enumerate}
\item (Sufficiency)\citep{SidiropoulosBro00TensorDecompUniqueness} The decomposition (\ref{eqn:R-CP-decomp}) is unique up to scaling and permutation if $\sum_{d=1}^D k_{\Bbf_d} \ge 2R + (D-1)$, where $k_{\Abf}$ is the $k$-rank of a matrix $\Abf$, i.e., the maximum value $k$ such that any $k$ columns are linearly independent.
\item (Necessity)\citep{LiuSidiropoulos01CRLB} If the decomposition (\ref{eqn:R-CP-decomp}) is unique up to scaling and permutation, then $\min_{d=1,\ldots,D} \mathrm{rank} (\Bbf_1 \odot \cdots \odot \Bbf_{d-1} \odot \Bbf_{d+1} \odot \cdots \odot \Bbf_D) = R$, which in turn implies that $\min_{d=1,\ldots,D} \left( \prod_{d' \ne d} \mathrm{rank}(\Bbf_{d'}) \right) \ge R$.
\item \citep{deLathauwer06CP} When $D=3$, $R \le p_3$ and $R(R-1) \le p_1(p_1-1)p_2(p_2-1)/2$, the decomposition (\ref{eqn:R-CP-decomp}) is unique for almost all such tensors except on a set of Lebesgue measure zero.
\item \citep{deLathauwer06CP} When $D=4$, $R \le p_4$ and $R(R-1) \le p_1p_2p_3(3p_1p_2p_3-p_1p_2-p_1p_3-p_2p_3-p_1-p_2-p_3+3)/4$, the decomposition (\ref{eqn:R-CP-decomp}) is unique for almost all such tensors except on a set of Lebesgue measure zero.
\end{enumerate}
\end{proposition}

Next we give a sufficient and necessary condition for local identifiability. The proof follows from a classical result \citep{Rothenberg71Identification} that relates local identifiability to the Fisher information matrix.
\begin{proposition}[Identifiability]
\label{prop:identifiability}
Given iid data points $\{(y_i,\xbf_i),i=1,\ldots,n\}$ from the tensor regression model. Let $\Bbf_0 \in \mathcal{\Bbf}$ be a parameter point and assume there exists an open neighborhood of $\Bbf_0$ in which the information matrix has a constant rank. Then $\Bbf_0$ is locally identifiable up to permutation if and only if
\begin{align*}
    I(\Bbf_0) = [\Jbf_1 \ldots \Jbf_D] \trans \left[ \sum_{i=1}^n \frac{\mu'(\eta_i)^2}{\sigma_i^2} (\vect \, \xbf_i) (\vect \, \xbf_i) \trans \right] [\Jbf_1 \ldots \Jbf_D]
\end{align*}
is nonsingular.
\end{proposition}

\smallskip
\noindent
\textit{Remark 3.1: }
Proposition \ref{prop:identifiability} explains the merit of tensor regression from another angle. For identifiability, the classical linear regression requires $\vect \, \xbf_i \in \real{\prod_d p_d}$, $i=1,\ldots,n$, to be linearly independent in order to estimate all parameters, which requires a sample size $n \ge \prod_d p_d$. The more parsimonious tensor regression only requires linearly independence of the ``collapsed" vectors $[\Jbf_1 \ldots \Jbf_D] \trans\vect \, \xbf_i \in \real{R(\sum_d p_d-D+1)}$, $i=1,\ldots,n$. The requirement on sample size is greatly lessened by imposing structure on the arrays.

\smallskip
\noindent
\textit{Remark 3.2: }
Although global identifiability is hard to check for a finite sample, a parameter point $\Bbf \in \mathcal{\Bbf}$ is asymptotically and
 globally identifiable as far as it admits a unique decomposition up to scaling and permutation and $\sum_{i=1}^n (\vect \, \xbf_i) (\vect \, \xbf_i) \trans$ has full rank for $n \ge n_0$, or, when considered stochastically, $\mathbf{E} [(\vect \, \Xbf) (\vect \, \Xbf) \trans]$ has full rank. To see this, whenever     $\sum_{i=1}^n (\vect \, \xbf_i) (\vect \, \xbf_i) \trans$ has full rank, the full coefficient array is globally identifiable and thus the decomposition is identifiable whenever it is unique.

Generalizing the concept of estimable functions for linear models, we call any linear combination of $\langle \xbf_i, \sum_{r=1}^R \betabf_1^{(r)} \circ \cdots \circ \betabf_D^{(r)}\rangle$, $i=1,\ldots,n$, as an estimable function. We can estimate estimable or collection of estimable functions even when the parameters are not identifiable.

\subsection{Asymptotics}
\label{sec:asymp}

The asymptotics for tensor regression follow from those for MLE or M-estimation. The key observation is that the nonlinear part of tensor model (\ref{eqn:r-tensorreg}) is a degree-$D$ polynomial of parameters. Then the classical Wald's continuity condition and Cram{\'e}r's smoothness condition become trivial. Our proofs in the Appendix are based on the uniform convergence conditions using Glivenko-Cantelli theory. For that purpose, note that the collection of polynomials $\{\langle \Bbf, \Xbf \rangle, \Bbf \in \mathcal{\Bbf}\}$ form a Vapnik-\u{C}ervonenkis (VC) class. Then standard theory for M-estimation \citep{vanderVaart98Asymp} applies.
\begin{theorem}[Consistency]
\label{thm:tensor-consistency}
Assume $\Bbf_0=\llbracket \Bbf_{01},\ldots,\Bbf_{0D}\rrbracket \in \mathcal{\Bbf}$ is (globally) identifiable up to permutation and  the array covariates $\Xbf_i$ are iid from a bounded  distribution. The MLE is consistent, i.e., $\hat \Bbf_n$ converges to $\Bbf_0$ (modulo permutation) in probability, in the following models: (1) normal tensor regression with a compact parameter space $\mathcal{\Bbf}_0 \subset \mathcal{\Bbf}$; (2) binary tensor regression; and (3) poisson tensor regression with a compact parameter space $\mathcal{\Bbf}_0 \subset \mathcal{\Bbf}$.
\end{theorem}
\noindent
\textit{Remark 4: }(Misspecified Rank)
In practice it is rare that the true regression coefficient $\Bbf_{\text{true}} \in \real{p_1 \times \cdots \times p_D}$ is exactly a low rank tensor. However the MLE of the rank-$R$ tensor model converges to the maximizer of function $M(\Bbf) = \mathbb{P}_{\Bbf_{\text{true}}} \ln p_{\Bbf}$ or equivalently $\mathbb{P}_{\Bbf_{\text{true}}} \ln (p_{\Bbf}/p_{\Bbf_{\text{true}}})$. In other words, the MLE is consistently estimating the best rank-$R$ approximation of $\Bbf_{\text{true}}$ in the sense of Kullback-Leibler distance.

To establish the asymptotic normality of $\hat \Bbf_n$, we note that  the log-likelihood function of   tensor regression model is
 quadratic mean differentiable (q.m.d).
 \begin{lemma}
\label{lemma:tensor-qmd}
Tensor regression model is quadratic mean differentiable (q.m.d.).
\end{lemma}
The next result follows from the asymptotic normality result for models that satisfy q.m.d. The proof is given in the Appendix.
\begin{theorem}[Asymptotic Normality]
\label{thm:tensor-normality}
For an interior point $\Bbf_0 = \llbracket \Bbf_{01}, \ldots, \Bbf_{0D} \rrbracket \in \mathcal{\Bbf}$ with nonsingular information matrix $\Ibf(\Bbf_{01}, \ldots, \Bbf_{0D})$ (\ref{eqn:fisher-info}) and $\hat \Bbf_n$ is consistent,
$$
\sqrt n [\vect (\hat \Bbf_{n1}, \ldots, \hat \Bbf_{nD}) - \vect (\Bbf_{01}, \ldots, \Bbf_{0D})]
$$
converges in distribution to a normal with mean zero and covariance $\Ibf^{-1}(\Bbf_{01}, \ldots, \Bbf_{0D})$.
\end{theorem}

\section{Regularized Estimation}
\label{sec:regularization}

The sample size in most neuroimaging studies is quite small, and thus even for a rank-1 tensor regression (\ref{eqn:model}), it is likely that the number of parameters exceeds  the sample size. Therefore, the $p>> n$ challenge is a rule rather than an exception in neuroimaging analysis, and regularization becomes essential. Even when the sample size exceeds the number of parameters, regularization is still useful for stabilizing the estimates and improving their risk property. We emphasize that there are a large number of regularization techniques for different purposes. Here we illustrate with using \emph{sparsity} regularization for identifying sub-regions that are associated with the response traits. This problem  can be viewed as an analogue of variable selection in the traditional vector-valued covariates. Toward that end, we maximize a regularized log-likelihood function
\begin{align*}
    \ell(\alpha,\gammabf,\Bbf_1,\ldots,\Bbf_D) - \sum_{d=1}^D \sum_{r=1}^R \sum_{i=1}^{p_d} P_\lambda(|\beta_{di}^{(r)}|,\rho),
\end{align*}
where $P_\lambda(|\beta|,\rho)$ is a scalar penalty function, $\rho$ is the penalty tuning parameter,  and $\lambda$ is an index for the penalty family. Some widely used penalties include: power family \citep{FrankFriedman93Bridge},  in which $P_\lambda(|\beta|,\rho) = \rho |\beta|^\lambda$, $\lambda \in (0,2]$, and in particular lasso \citep{Tibshirani96Lasso} ($\lambda=1$) and ridge ($\lambda=2$); elastic net \citep{ZouHastie05Enet}, in which $P_\lambda(|\beta|, \rho) = \rho [(\lambda-1) \beta^2/2 + (2-\lambda) |\beta|], \lambda \in [1,2]$; and SCAD \citep{FanLi01SCAD}, in which $\partial / \partial |\beta| P_{\lambda}(|\beta|, \rho) = \rho \left\{ 1_{\{|\beta| \le \rho\}} + (\lambda \rho - |\beta|)_+ /(\lambda-1)\rho 1_{\{|\beta| > \rho\}} \right\}$, $\lambda>2$, among many others.
Choice of penalty function and tuning parameters $\rho$ and $\lambda$ depends on particular purposes: prediction, unbiased estimation, or region selection.

Regularized estimation for tensor models incurs slight changes in Algorithm \ref{algo:br-algo}. When updating $\Bbf_d$, we simply fit a penalized GLM regression problem,
\begin{align*}
    & \Bbf_d^{(t+1)} = & \mbox{argmax}_{\Bbf_d} \, \ell(\alpha^{(t)},\gammabf^{(t)}, \Bbf_1^{(t+1)}, \ldots, \Bbf_{d-1}^{(t+1)}, \Bbf_d, \Bbf_{d+1}^{(t)}, \ldots, \Bbf_D^{(t)}) - \sum_{r=1}^R \sum_{i=1}^{p_d} P_\lambda(|\beta_{di}^{(r)}|,\rho),
\end{align*}
for which many software packages exist. Same paradigm certainly applies to regularizations other than sparsity. The fitting procedure boils down to alternating regularized GLM regression. The monotone ascent property of Algorithm \ref{algo:br-algo} is retained under the modified algorithm, giving rise to stability in the estimation algorithm. Convex penalties, such as elastic net and power family with $\lambda \ge 1$, tend to convexify the objective function and alleviate the local maximum problem. On the other hand, concave penalty such as power family with $\lambda<1$ and SCAD produces more unbiased estimates but the regularized objective function is more ruggy and in practice the algorithm should be initialized from multiple start points to increase the chance of finding a global maximum. Many methods are available to guide the choice of the tuning parameter $\rho$ and/or $\lambda$ for regularized GLM, notably AIC, BIC and cross validation. For instance the recent work \citep{ZhouArmagan11SparsePath} derives BIC type criterion for GLM with possibly non-concave penalties such as power family, which can be applied to regularized tensor regression models in a straightforward way.

Two remarks are in order. First, it is conceptually possible to apply these regularization techniques directly to the full coefficient array $\Bbf \in \real{\prod_d p_d}$ without considering any structured decomposition as in our models. That is, one simply treats $\vect \Xbf$ as the predictor vector as employed in the classical total variation regularization in image denoising and recovery. However, for the brain imaging data, we should bear in mind the dimensionality of the imaging arrays. For instance, to the best of our knowledge, no software is able to deal with fused lasso or even simple lasso on $64^3 = 262,144$ or $256^3 = 16,777,216$ variables. This ultrahigh dimensionality certainly corrupts the statistical properties of the regularized estimates too. Second, penalization is only one form of regularization. In specific applications, prior knowledge often suggests various constraints among parameters, which may be exploited to regularize parameter estimate. For instance, for MRI imaging data, sometimes it may be reasonable to impose symmetry on the parameters along the coronal plane, which effectively reduces the dimensionality by $p_dR/2$. In many applications, nonnegativity of parameter values is also enforced.

\section{Numerical Analysis}
\label{sec:numerics}

We have carried out an extensive numerical study to investigate the finite sample performance of the proposed methods. In this section,  we report some selected results from some synthetic examples and an analysis of a real brain imaging data.

\subsection{2D Shape Examples}

We first elaborate on the illustrative example given in Section \ref{sec:basic-model} with a collection of 2D shapes. In the first study, we varied the sample size to demonstrate the consistency of tensor regression estimation. In the second study, we illustrate regularized tensor regression estimation.

\subsubsection{Tensor Regression Estimation}

We employ the tensor model setup in Section \ref{sec:basic-model}, where the response is normally distributed with mean, $\eta = \gamma\trans \Zbf + \langle \Bbf,\Xbf \rangle$, and standard deviation one. Here $\Xbf$ is a $64 \times 64$ 2D matrix, $\Zbf$ is a $5$-dimensional covariate vector, both of which have standard normal entries, $\gammabf = (1, 1, 1, 1, 1)\trans$, and $\Bbf$ is binary with the true signal region equal to one and the rest zero. We examine various sample sizes at $n=500, 750$ and $1000$, and report the results under a tensor model whose rank is determined by BIC. Table~\ref{tab:ex1matrix} summarizes the mean results out of 100 data replications. Reported criterion is root mean squared error (RMSE) for both the regular vector coefficient $\gammabf$ and the array coefficient $\Bbf$. It is clearly seen that the estimation accuracy increases along with the sample size, demonstrating the consistency of the proposed method.

\begin{table}[t]
\caption{Tensor regression estimation for the 2D shape examples. Reported are mean RMSE and its standard deviation (in parenthesis) based on 100 data replications.}
\begin{center}
\begin{tabular}{lcccc}
\toprule
Shape      &  Param.  & $n=500$ & $n=750$ & $n=1000$ \\ \midrule
Square     & $\gammabf$ & 0.0486 (0.0173) & 0.0387 (0.0129) & 0.0325 (0.0104) \\	
                  & $\Bbf$            & 0.0091 (0.0007) & 0.0071 (0.0005) & 0.0059 (0.0004) \\ \hline	
T-shape   & $\gammabf$ & 0.0612 (0.0209) & 0.0423 (0.0135) & 0.0356 (0.0112) \\	
                  & $\Bbf$            & 0.0160 (0.0010) & 0.0113 (0.0006) & 0.0091 (0.0005) \\ \hline	
Cross       & $\gammabf$ & 0.0610 (0.0199) & 0.0425 (0.0140) & 0.0345 (0.0104) \\	
                  & $\Bbf$            & 0.0159 (0.0011) & 0.0112 (0.0006) & 0.0090 (0.0005) \\ \hline	
Disk          & $\gammabf$ & 0.5702 (0.1793) & 0.1804 (0.0588) & 0.1263 (0.0402) \\	
                  & $\Bbf$            & 0.2125 (0.0218) & 0.0765 (0.0050) & 0.0622 (0.0017) \\ \hline	
Triangle   & $\gammabf$ & 0.6327 (0.2337) & 0.2111 (0.0752) & 0.1491 (0.0541) \\	
                  & $\Bbf$            & 0.2343 (0.0254) & 0.0992 (0.0066) & 0.0775 (0.0021) \\ \hline	
Butterfly   & $\gammabf$ & 1.4385 (0.5561) & 0.5870 (0.1639) & 0.3884 (0.1310) \\	
                  & $\Bbf$            & 0.5536 (0.0570) & 0.2669 (0.0193) & 0.1998 (0.0071) \\ 	
\bottomrule
\end{tabular}
\label{tab:ex1matrix}
\end{center}
\end{table}

\subsubsection{Regularized Tensor Regression Estimation}

\begin{figure}
\begin{center}
\begin{tabular}{cc}
\includegraphics[width=4.4in]{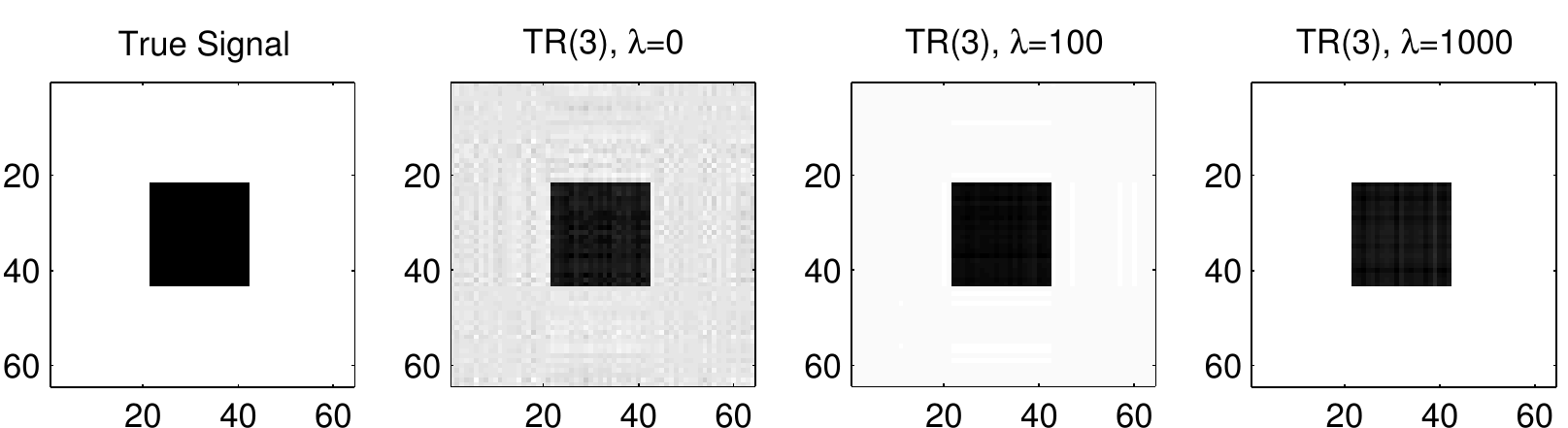} \\ \includegraphics[width=4.4in]{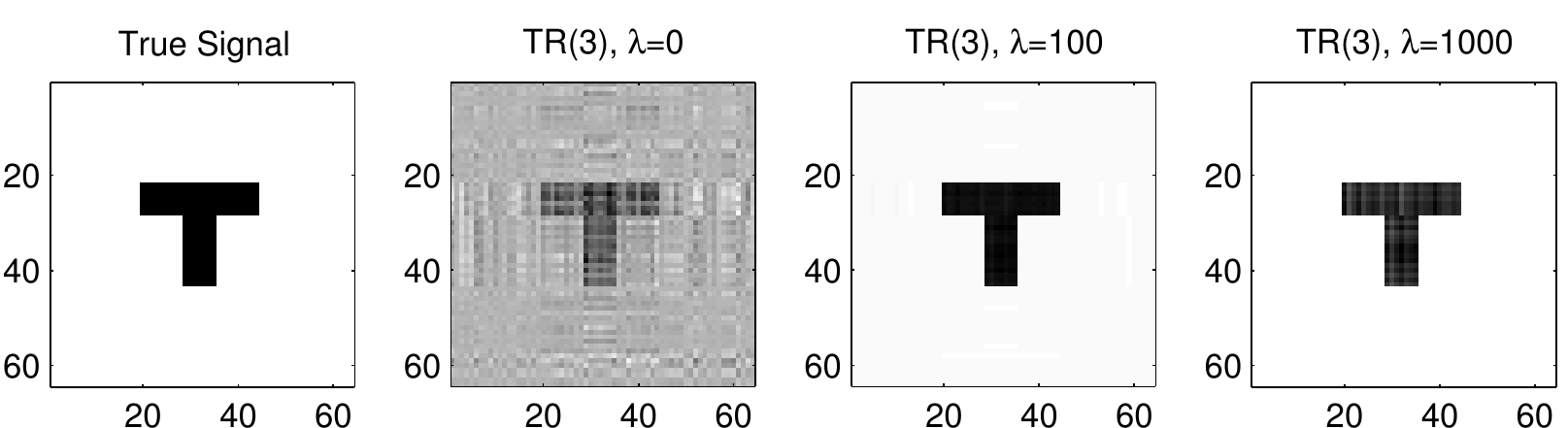}    \\
\includegraphics[width=4.4in]{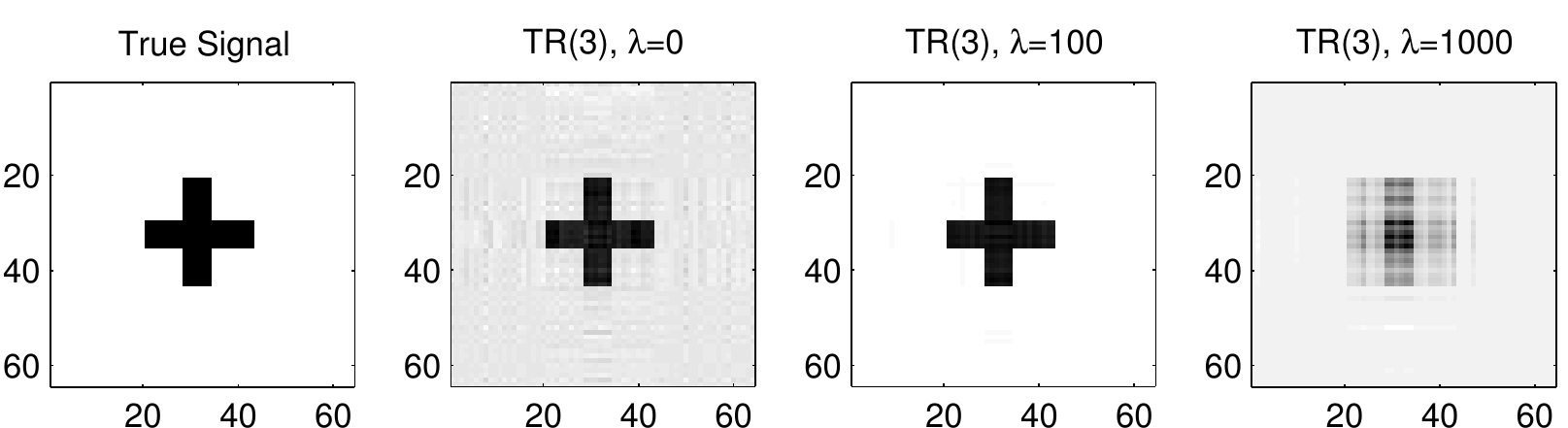} \\ \includegraphics[width=4.4in]{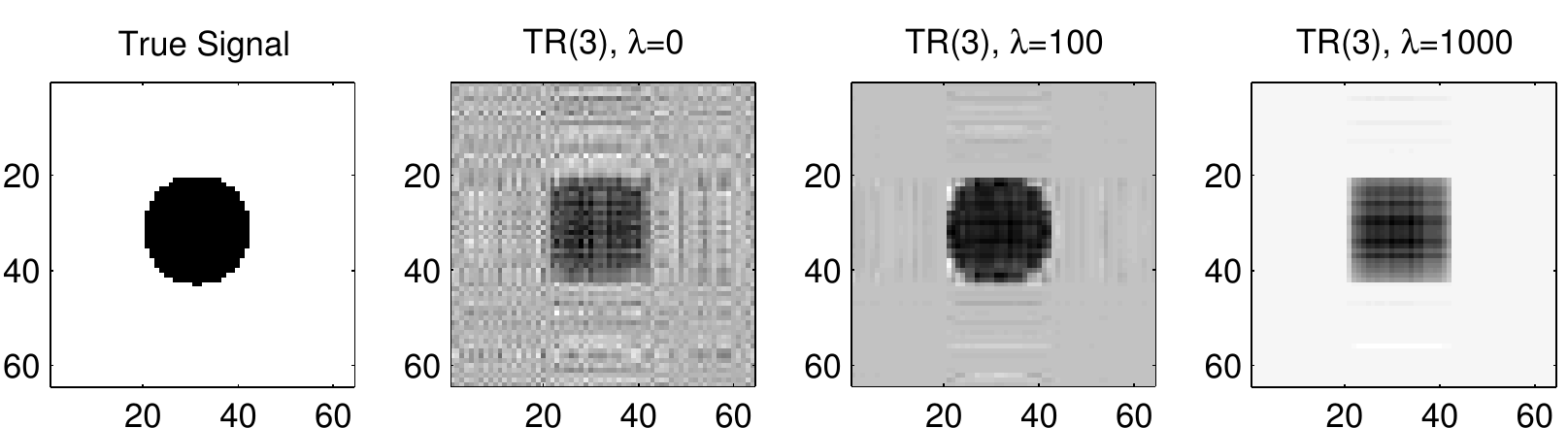}    \\
\includegraphics[width=4.4in]{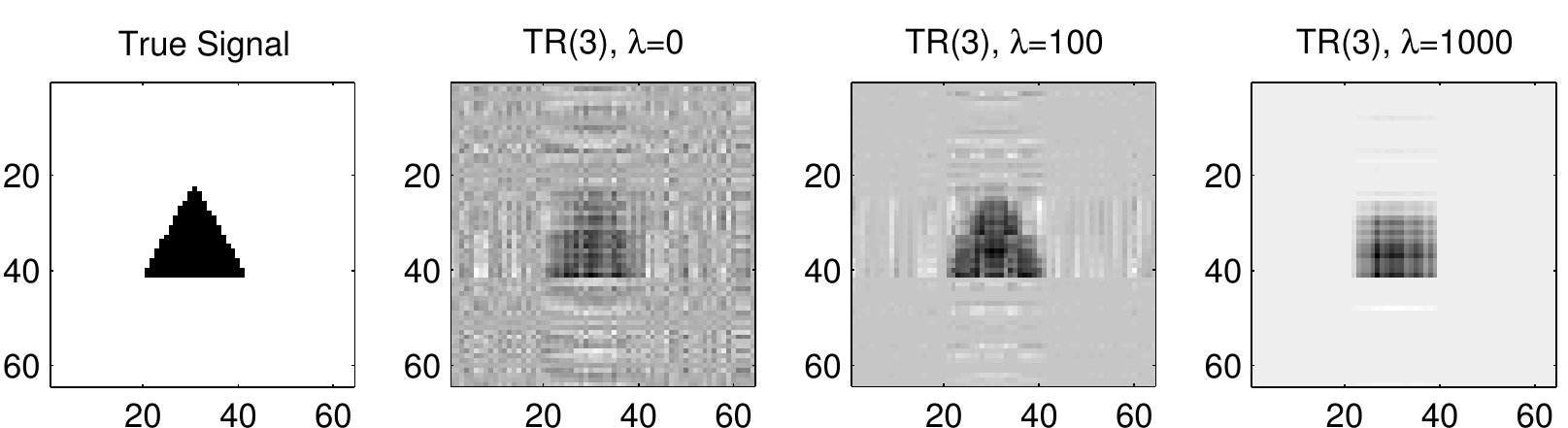} \\ \includegraphics[width=4.4in]{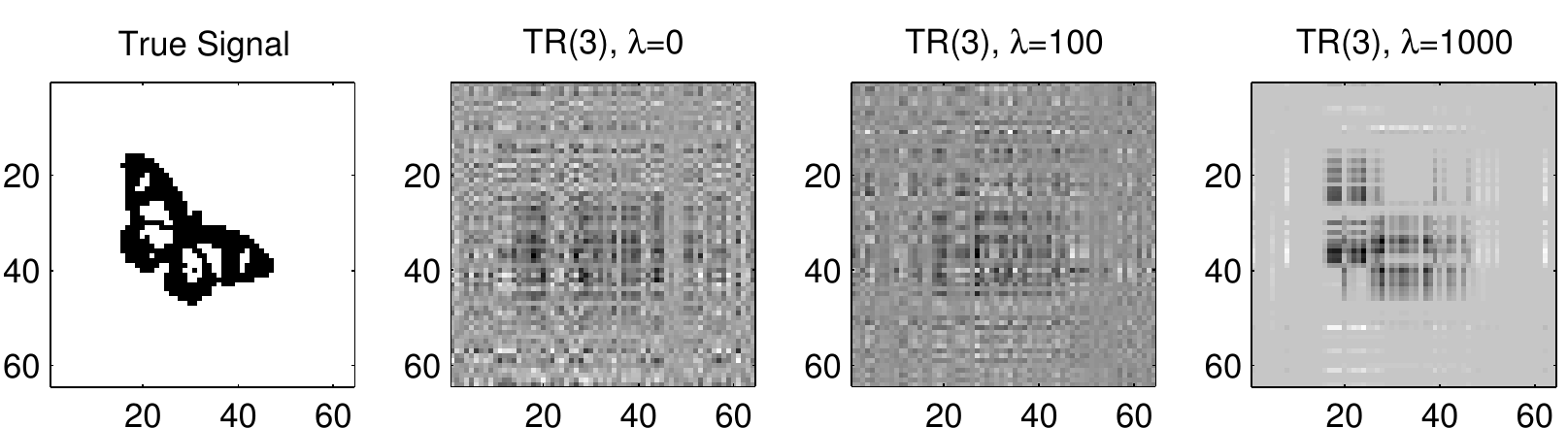}
\end{tabular}
\caption{Demonstration of lasso regularization. The matrix variate has size 64 by 64 with entries generated as independent standard normals. The regression coefficient for each entry is either 0 (white) or 1 (black). The sample size is 500.}\label{fig:shapes-lasso}
\end{center}
\end{figure}

Next we revisit the example in Section \ref{sec:basic-model} to illustrate regularized tensor regression estimation. The setup is the same as that in Figure \ref{fig:shapes} except that the sample size is reduced to 500, which is only barely larger than the number of parameters $417 = 5+3 \times (64+64)$ of a rank-3 tensor model. Figure \ref{fig:shapes-lasso} shows the outcome of applying the lasso penalty to $\Bbf_d$ in the rank-3 tensor regression model. Recovered signals at three different values of $\lambda = 0, 100, 1000$ are displayed. Without regularization ($\lambda = 0$), the rank-3 tensor regression is difficult to recover some signals such as triangle, disk and butterfly, mainly due to a very small sample size. On the other hand, excessive penalization compromises the quality of recovered signals too, as evidently in those shapes at $\lambda = 1,000$. Regularized estimation with an appropriate amount of shrinkage improves estimation quality, as seen in the triangle and disk at $\lambda=100$ and in butterfly at $\lambda=1000$. In practice the tuning parameter is chosen by a certain model selection criterion such as BIC or cross validation. Moreover, we have experimented with the bridge and SCAD penalties for the same data and obtained similar results. The desirable unbiased (or nearly unbiased) estimates from those concave penalties are reflected by the improved contrast in the recovered signal. But for the sake of space, we do not show those figures here.

\subsection{Attention Deficit Hyperactivity Disorder Data Analysis}

We applied our methods to the attention deficit hyperactivity disorder (ADHD) data from the ADHD-200 Sample Inititive (\textsf{http://fcon$\underline{\mbox{ }}$1000.projects.nitrc.org/indi/adhd200/}). ADHD is a common childhood disorder and can continue through adolescence and adulthood. Symptoms include difficulty in staying focused and paying attention, difficulty in controlling behavior, and over-activity. The data set that we used is part of the ADHD-200 Global Competition data sets. It consists of $776$ subjects, with 491 normal controls and 285 combined ADHD subjects. Among them, there are $442$ males whose mean age is $11.9815$ years with standard deviation $3.1355$ years, and $287$ females whose mean age is $11.8617$ years with standard deviation $3.4875$ years. We removed 47 subjects due to the missing observations or poor image quality. Rs-fMRIs and T1-weighted images were acquired for each subject.
The T1-weighted images were  preprocessed  by standard steps including AC (anterior commissure) and PC (posterior commissure) correction,  ÒN2Ó bias field correction,  skull-stripping, intensity inhomogeneity correction,   cerebellum removal,  segmentation, and registration. After segmentation,  the brains were segmented into four different tissues: grey matter (GM), white matter (WM), ventricle (VN), and cerebrospinal fluid (CSF). We quantified the local volumetric group differences by generating RAVENS maps \citep{Davatzikos2001} for the whole brain and each of the segmented tissue type (GM, WM, VN, and CSF) respectively, using the deformation field we obtained during registration. RAVENS methodology is based on a volume-preserving spatial transformation, which ensures that no volumetric information is lost during the process of spatial normalization, since this process changes an individual's brain morphology to conform it to the morphology of a template. In addition to image covariates,  we include the subjects' age, gender, and whole brain volume as regular covariates. One scientific question of interest is to understand association between the disease outcome and the brain image patterns after adjustment for the clinical and demographical variables. First, we examined the case with real image covariates and simulated responses. Our goal is to study the empirical performance of our methods under various response models. Secondly, we showed the performance of the regularized estimation in terms of region selection. Finally, we applied the method to the data with the true observed binary response.

\subsubsection{Real Image Covariates and Simulated Response}
\label{sec:adhdsim}

We first consider a number of GLMs with the real brain image covariates, where $\eta = \gammabf\trans \Zbf + \langle \Bbf,\Xbf \rangle$, the signal tensor $\Bbf$ admits a certain structure, $\gammabf = (1, 1, 1)\trans$, $\Xbf$ denotes the 3D MRI image with dimension $256 \times 256 \times 198$, and $\Zbf$ denotes the vector of age, gender and whole brain volume. We consider two structures for $\Bbf$. The first admits a rank one decomposition, with $\Bbf_1 \in \real{256 \times 1}$,  $\Bbf_2 \in \real{256 \times 1}$, and $\Bbf_3 \in \real{198 \times 1}$, and all of whose $(90 + j)$th element equal to $\sin(j \pi/14)$ for $j=0, 1, \ldots, 14$. This corresponds to a single-ball signal in a 3D space. The second admits a rank two decomposition, with $\Bbf_1 \in \real{256 \times 2}$,  $\Bbf_2 \in \real{256 \times 2}$, and $\Bbf_3 \in \real{198 \times 2}$. All the first columns of $\Bbf_d$ have their $(90 + j)$th element equal to $\sin(j \pi/14)$, and the second columns of $\Bbf_d$ have their $(140 + j)$th element equal to $\sin(j \pi/14)$ for $j=0, 1, \ldots, 14$. This mimics a two-ball signal in the 3D space. We then generate the response through the GLM models: for the normal model, $Y \sim \mathrm{Normal}(\mu, 1)$, where $\mu = \eta$; for the binomial model, $Y \sim \mathrm{Bernoulli}(p)$, with $p = 1 / [1 + \exp(-0.1 \eta)]$; and for the poisson model, $Y \sim \mathrm{Poission}(\mu)$, with $\mu = \exp(0.01 \eta)$. Table~\ref{tab:ex2adhd} summarizes the average RMSE and its standard deviation out of 100 data replications. We see that the normal and poisson responses both have competitive performance, whereas the binomial case is relatively more challenging. The two-ball signal is more challenging than a one-ball signal, and overall the tensor models work well across different response types and different signals.

\begin{table}[t]
\caption{Tensor regression estimation for the ADHD data. Reported are mean RMSE and its standard deviation (in parenthesis) of evaluation criteria based on 100 data replications.}
\begin{center}
\begin{tabular}{lcccc}
\toprule
Signal     & Param.     & Normal                & Binomial              & Poisson \\ \midrule
one-ball  & $\gammabf$ & 0.0639 (0.0290) & 0.2116 (0.0959) & 0.0577 (0.0305) \\
                 & $\Bbf$            & 0.0039 (0.0002) & 0.0065 (0.0002) & 0.0064 (0.0002) \\ \hline
two-ball   & $\gammabf$ & 0.0711 (0.0310) & 0.3119 (0.1586) & 0.0711 (0.0307) \\
                 & $\Bbf$            & 0.0058 (0.0002) & 0.0082 (0.0003) & 0.0083 (0.0003) \\
\bottomrule
\end{tabular}
\label{tab:ex2adhd}
\end{center}
\end{table}

\subsubsection{Regularized Estimation}

\begin{figure}
\begin{center}
\begin{tabular}{cc}
\vspace{-0.33in}
\small{True Signal (a)} & \small{(b)} \\
\vspace{-0.2in}
\includegraphics[width=2.5in,height=2.9in]{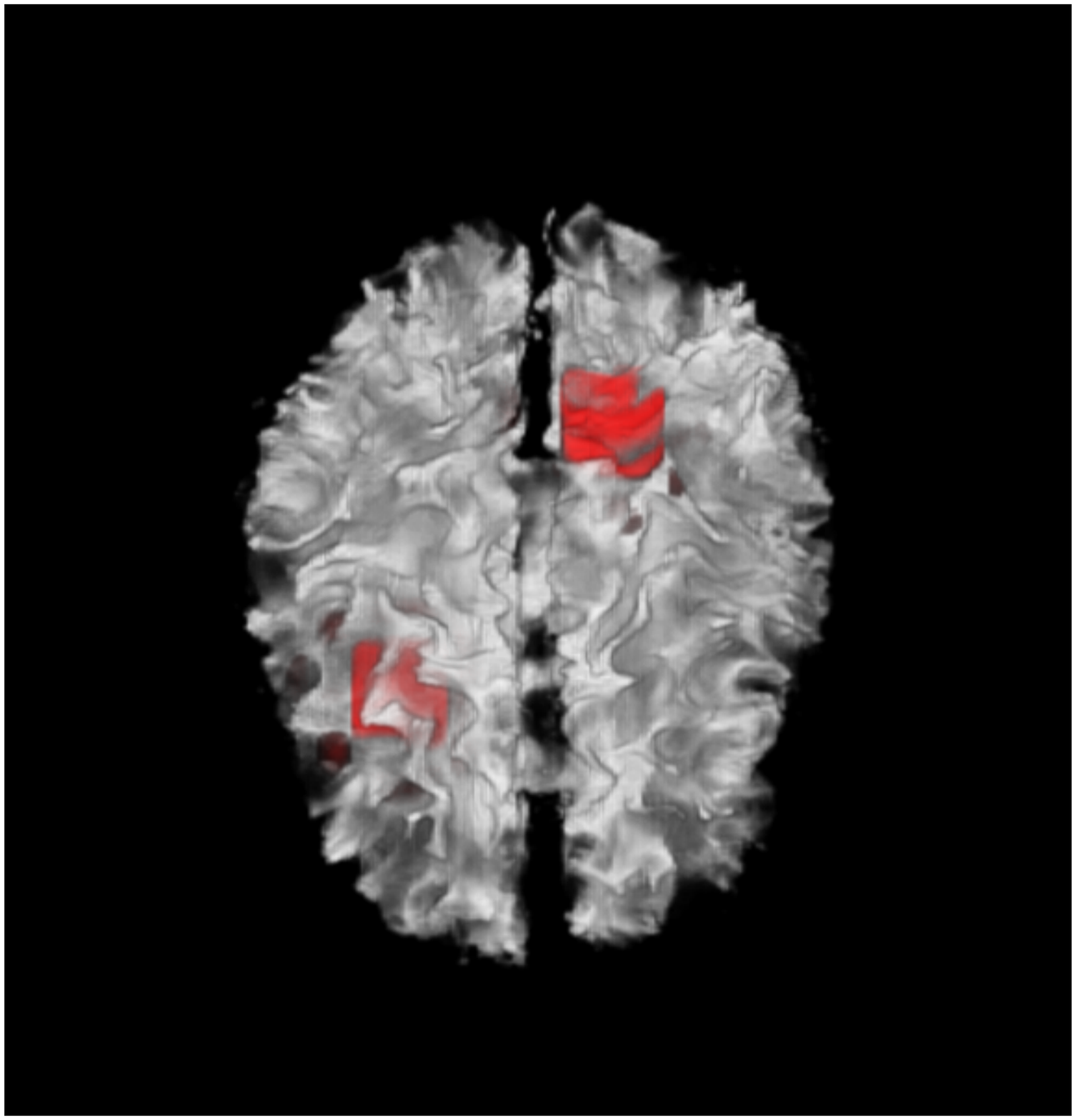} & \hspace{-0.35in} \includegraphics[width=2.5in,height=2.9in]{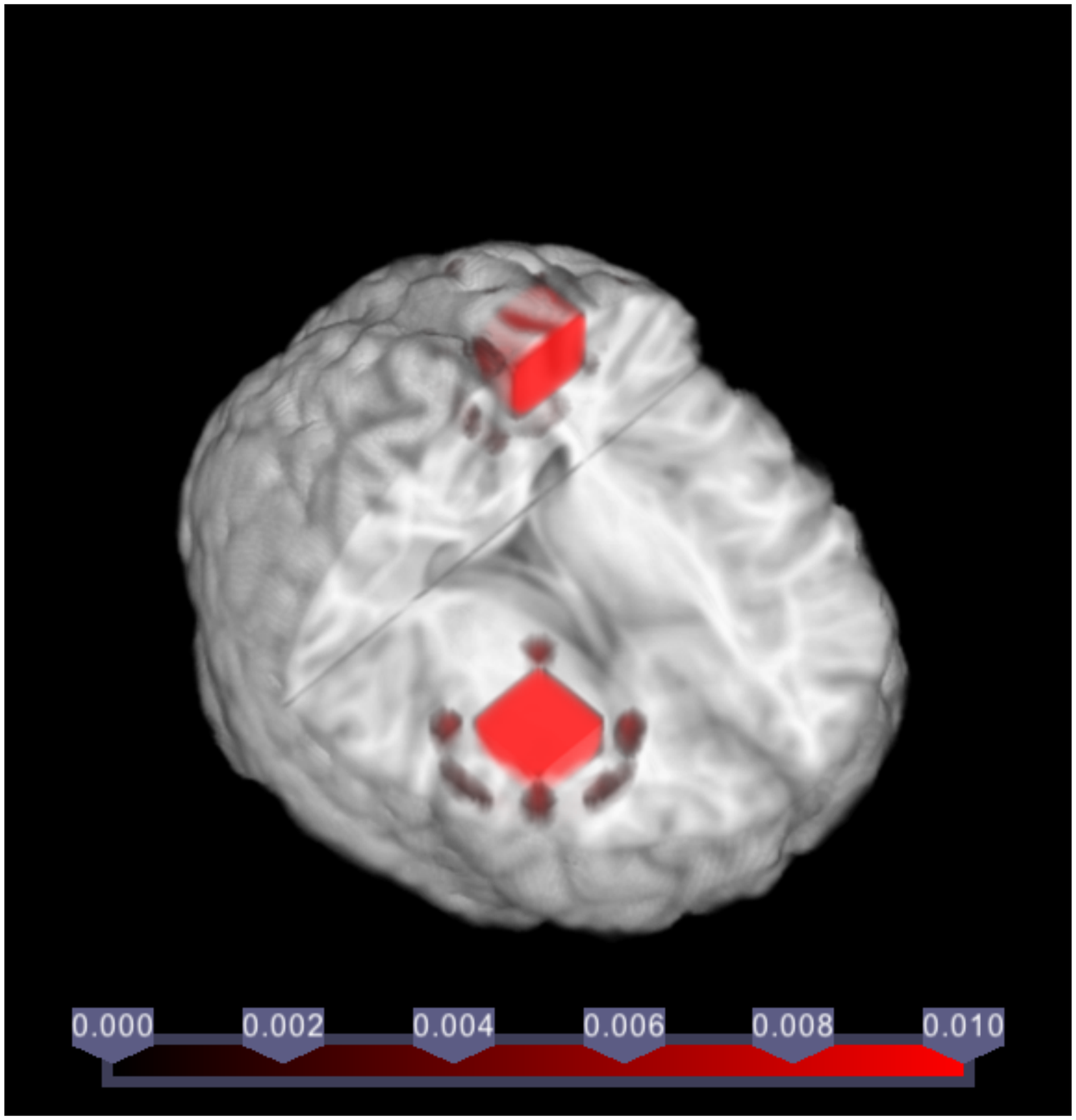} \\
\vspace{-0.33in}
\small{Unpenalized estimation (a)} & \small{(b)} \\
\vspace{-0.2in}
\includegraphics[width=2.5in,height=2.9in]{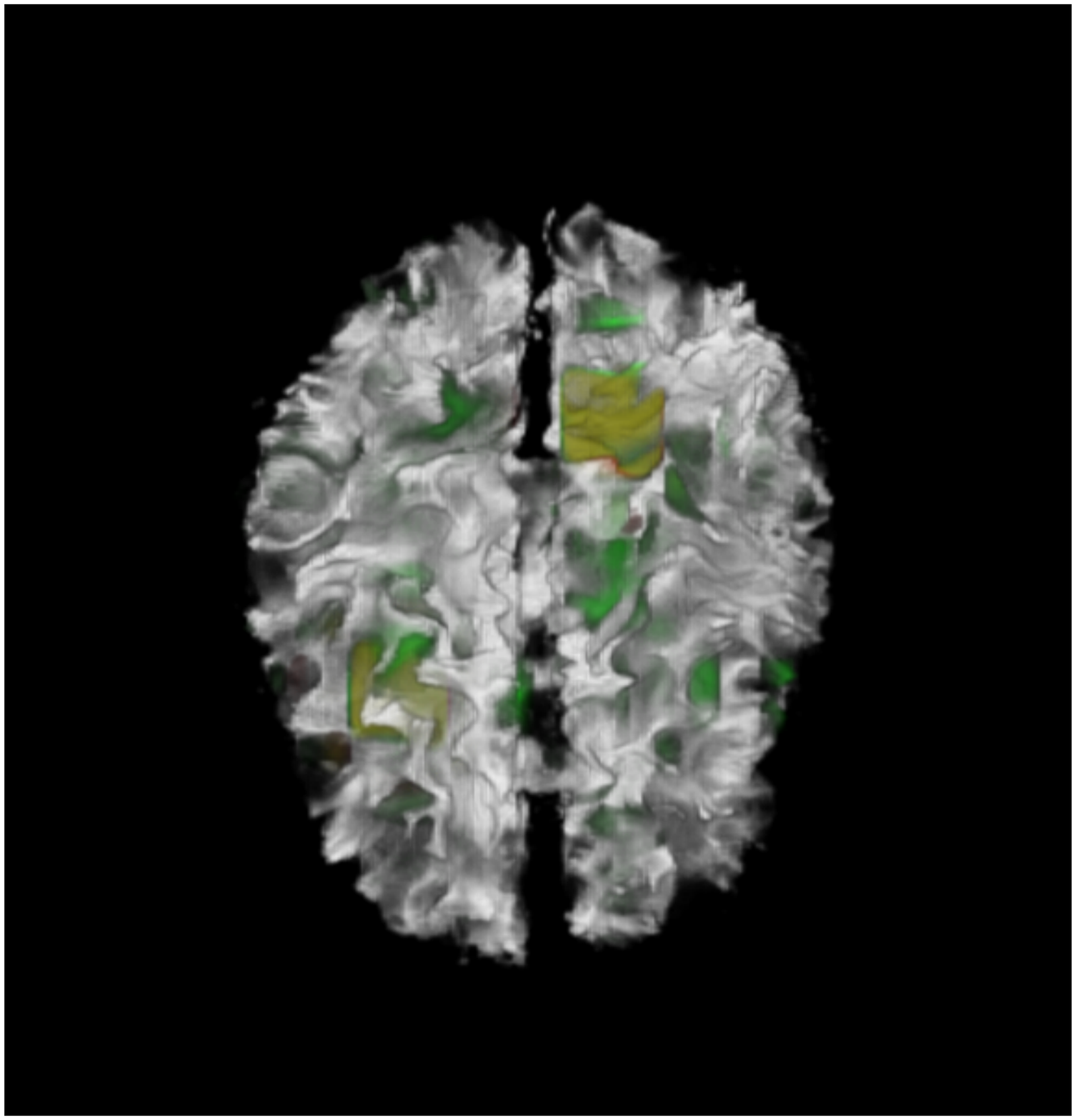} & \hspace{-0.35in} \includegraphics[width=2.5in,height=2.9in]{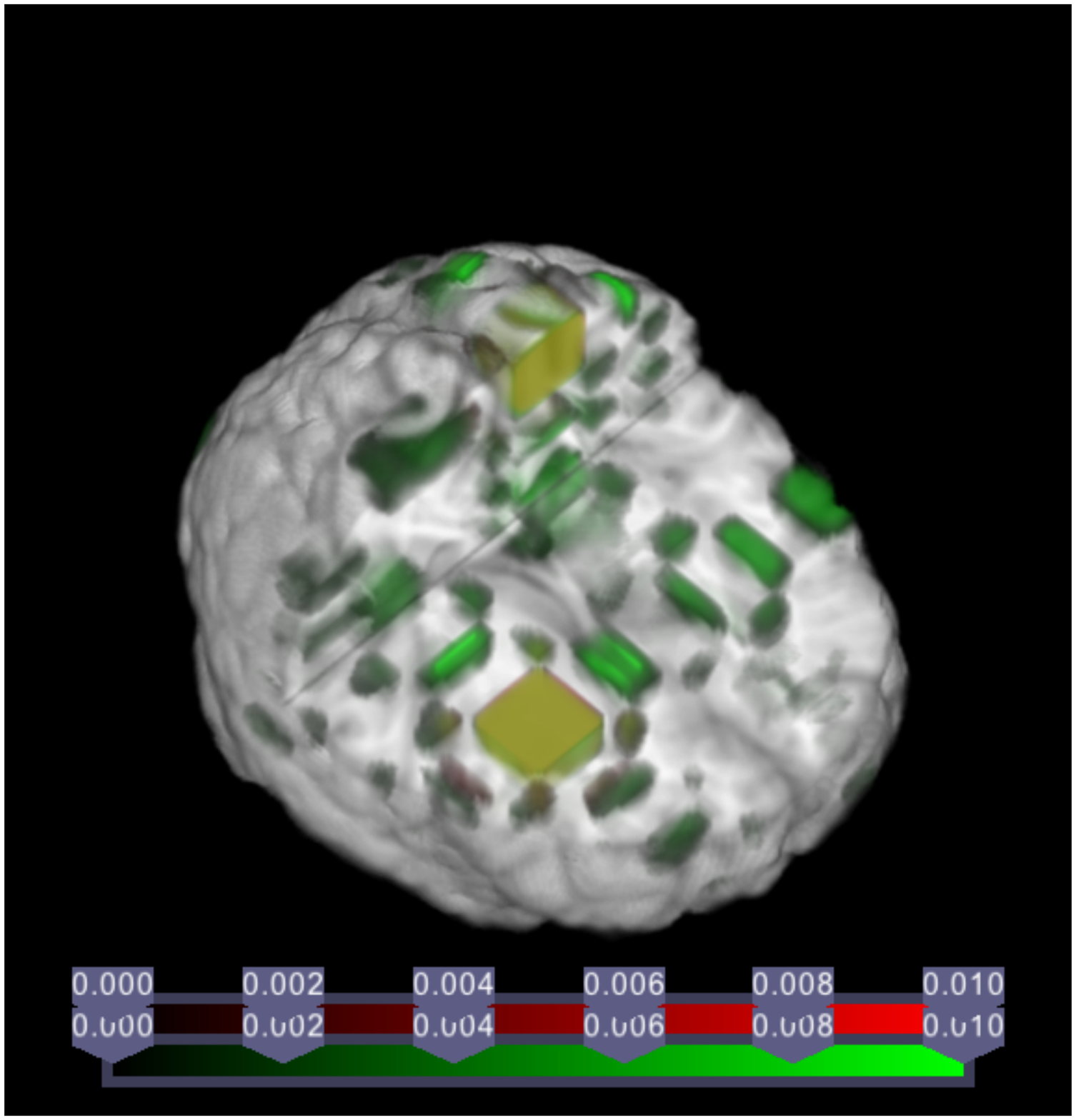} \\
\vspace{-0.33in}
\small{Lasso estimation (a)} & \small{(b)} \\
\vspace{-0.2in}
\includegraphics[width=2.5in,height=2.9in]{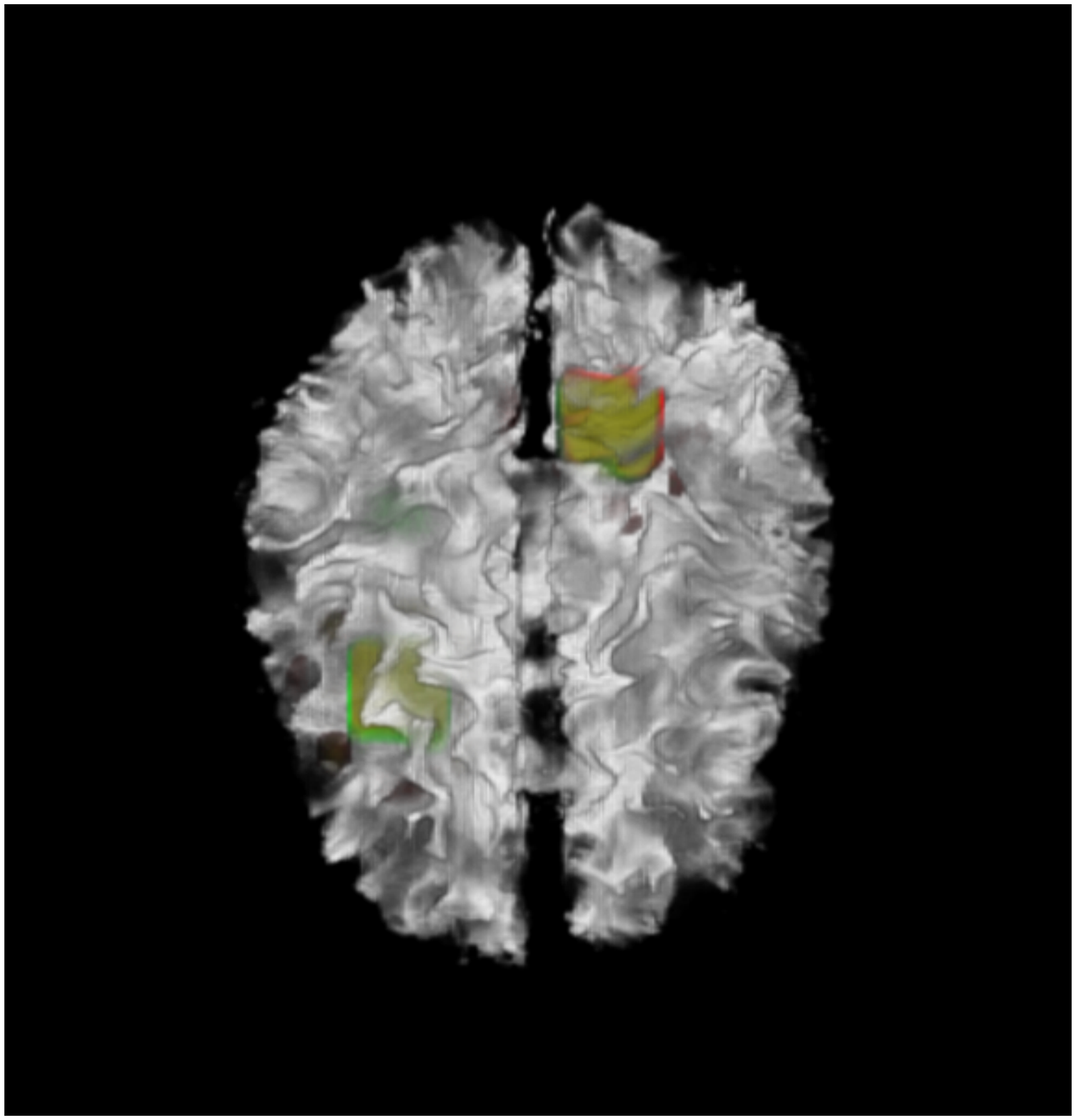} & \hspace{-0.35in} \includegraphics[width=2.5in,height=2.9in]{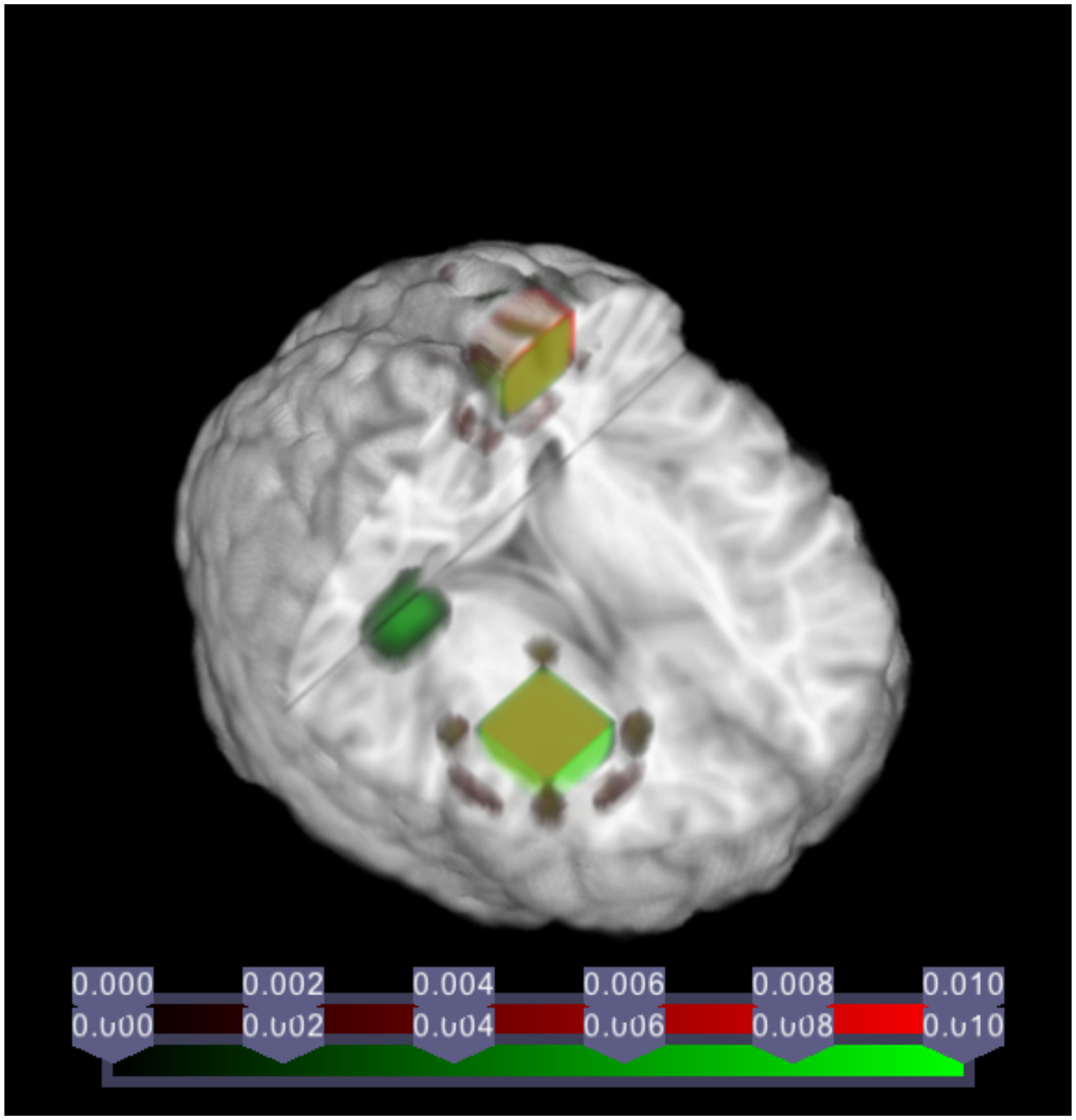} \\
\end{tabular}
\caption{Region selection. The true signal regions are colored in red, the estimated signal regions are in green, and the overlapped regions are in yellow. The left panel is the true or estimated signal overlaid on a randomly selected subject, and the right panel is a 3D rendering of the true or estimated signal overlaid on the template}\label{fig:adhdreg}
\end{center}
\end{figure}

Next we focus on the ability of the proposed regularized tensor regression model to identify relevant regions in brain associated with the response. This problem is an analogue of variable selection in the traditional regression with vector-valued covariates. We employ the two-ball signal and the normal model in Section \ref{sec:adhdsim}. Figure \ref{fig:adhdreg} shows images with the true signal, the un-regularized tensor regression estimate, and the regularized tensor regression estimates with a lasso penalty, respectively, overlaid on an image of an arbitrarily chosen subject, or on a 3D rendering of a template. The plots clearly show that the true sparse signal regions can be well recovered through regularization.

\subsubsection{Real Data Analysis}

\begin{figure}
\begin{center}
\includegraphics[height=1.85in]{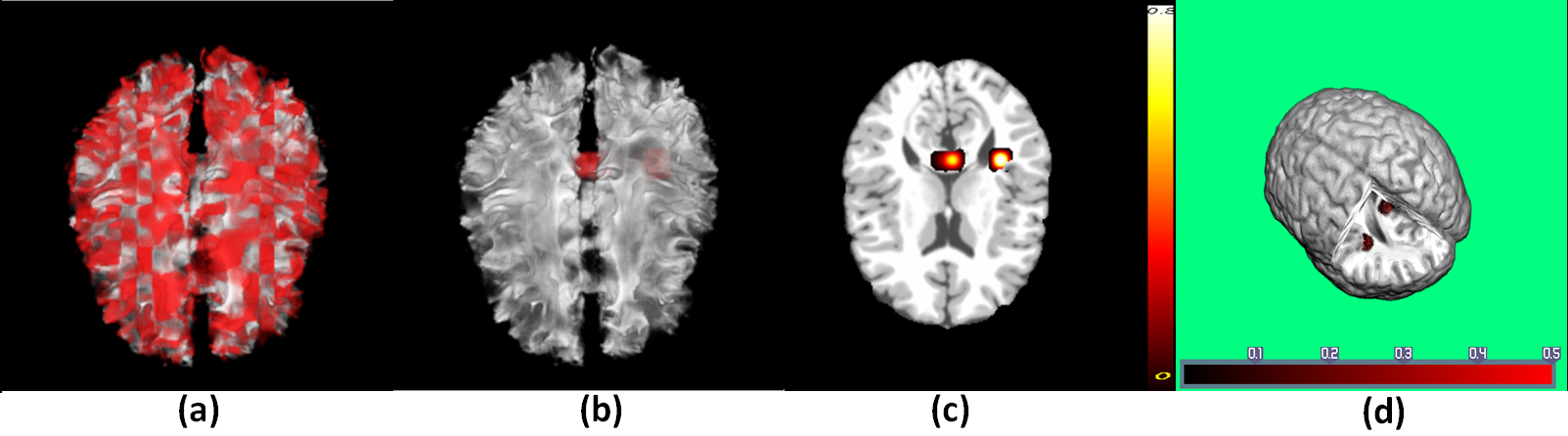}
\caption{Application to the ADHD data. Panel (a) is  the unpenalized estimate overlaid on a randomly selected subject; (b) is the regularized estimate overlaid on a randomly selected subject; (c) is a selected slice of the regularized estimate overlaid on the template; and (d) is a 3D rendering of the regularized estimate overlaid on the template.}\label{fig:adhd}
\end{center}
\end{figure}

Finally, we analyze the ADHD data with the observed binary diagnosis status as the response. We fitted a rank-3 tensor logistic regression model, since in practice it is rare that the true signal would follow an exact reduced rank formulation. We also applied the regularized estimation using a lasso penalty. Figure \ref{fig:adhd} shows      the results. Inspecting Figure ~\ref{fig:adhd} reveals two regions of interest: left temporal lobe white matter and the splenium that connects parietal and occipital cortices across the midline in the corpus callosum. The anatomical disturbance in the temporal lobe has been consistently revealed and its interpretation would be consistent with a finer-grained analysis of the morphological features of the cortical surface, which reported prominent volume reductions in the temporal and frontal cortices in children with ADHD compared with matched controls \citep{Sowell2003}.  Moreover, a  reduced size of the splenium is the most reliable finding in the corpus callosum \citep{Valera2007}.

\section{Discussion}
\label{sec:discussion}

We have proposed a tensor decomposition based approach for regression modeling with array covariates. The curse of dimensionality is circumvented by imposing a low rank approximation to the extremely high-dimensional full coefficient array. This allows development of a fast estimation algorithm and regularization. Numerical analysis demonstrates that, despite its massive reduction, the method works very well in recovering various geometric as well as natural shape images.

We view the method of this article a first step toward a more general area of array regression analysis, and the idea can be extended to a wide range of problems. We describe a few potential future directions here. First, although we only present results for models with a conventional covariate vector and an array covariate, the framework applies to arbitrary combination of array covariates. This provides a promising approach to the analysis of multi-modality data which becomes increasingly available in modern neuroimaging and medical studies. Besides the main effects of different array covariates, the interaction between them will be of interest, and can be studied under this framework too. Second, we remark that our modeling approach and algorithm equally apply to many general loss functions occurring in classification and prediction. For example, for a binary response $Y \in \{0,1\}$, the hinge loss takes the form
\begin{eqnarray*}
    \sum_{i=1}^{n}[1 - y_i \{\alpha + \gammabf \trans \zbf_i + \langle \sum_{r=1}^R \betabf_1^{(r)} \circ \betabf_2^{(r)} \circ \cdots \circ \betabf_D^{(r)},\xbf_i \rangle\}]_+
\end{eqnarray*}
and should play an important role in support vector machines with array variates. Third, in this article rotation has not been explicitly considered in the modeling. When prior knowledge indicates, sometimes it is prudent to work in polar coordinates. For example, the `disk' signal in Figure \ref{fig:shapes} can be effectively captured by a rank-1 outer product if the image is coded in polar coordinates. A diagonal signal array has full rank and cannot be approximated by any lower rank array, but if changed to polar coordinates, the rank reduces to one. Some of these extensions are currently under investigation. In summary,  we believe that the proposed methodology timely answers calls in modern neuroimaging data analysis, whereas the general methodology of tensor regression is to play a useful role and also deserves more attention in statistical analysis of high-dimensional complex imaging data.

\baselineskip=13pt
\bibliography{ref-tensor}
\bibliographystyle{Chicago}

\end{document}